\documentclass[aps,prx,10pt,superscriptaddress, noeprint, twocolumn, floatfix]{revtex4-2}

\usepackage{bbold}
\usepackage{comment}
\usepackage[colorlinks, citecolor=red]{hyperref}
\usepackage[utf8]{inputenc}
\usepackage[T1]{fontenc}    %
\usepackage[english]{babel}
\usepackage{graphicx}
\usepackage{amsmath,amssymb,amsthm,bbm, mathtools} %
\usepackage{mathrsfs}
\usepackage{xcolor}
\usepackage[normalem]{ulem}
\usepackage{placeins}
\usepackage{braket}
\usepackage{cancel}

\usepackage{comment}

\DeclareMathSymbol{\shortminus}{\mathbin}{AMSa}{"39}
\renewcommand{\vec}{\boldsymbol}

\begin{document}

\title{Superconducting singlet-triplet qubits}
\author{Anatoliy Lotkov}
\affiliation{Department of Physics, University of Basel, Klingelbergstrasse 82, 4056 Basel, Switzerland}
\author{Maria Spethmann}
\affiliation{Department of Physics, University of Basel, Klingelbergstrasse 82, 4056 Basel, Switzerland}
\author{Daniel Loss}
\affiliation{Department of Physics, University of Basel, Klingelbergstrasse 82, 4056 Basel, Switzerland}
\affiliation{Physics Department, King Fahd University of Petroleum and Minerals, 31261, Dhahran, Saudi Arabia}
\affiliation{Quantum Center, KFUPM, 31261, Dhahran, Saudi Arabia}
\affiliation{RDIA Chair in Quantum Computing, 31261, Dhahran, Saudi Arabia}

\begin{abstract}
Hybrid devices integrating quantum dots with Josephson junctions are gaining interest because they combine spin-based quantum computing with circuit quantum electrodynamics (circuit QED) methods. In particular, Andreev spin qubits have shown significant experimental progress including strong two-qubit coupling, and are predicted to exhibit all-to-all connectivity. Here we propose superconducting singlet-triplet (SST) qubits that rely on parallel-aligned double quantum dots in Josephson junctions. While Andreev spin qubits require spin-orbit interaction to unlock the spin degree-of-freedom, SST qubits do not require spin-orbit interaction, making the advantages of hybrid devices available to a wider range of materials. Similar to Andreev spin qubits, the qubit states couple to the superconducting phase across the junction, which allows for control and readout using circuit QED, and supports all-to-all connectivity.  Only $N$ flux lines are required to perform any single- and two-qubit gate among $N$ qubits, and thus the overhead of control lines is small. Finally, linear protection from charge or flux noise makes these qubits interesting candidates for a future quantum processor. 
\end{abstract}

\maketitle

\section{Introduction}

Promising platforms for quantum computing include superconducting qubits \cite{Devoret1995,Vool2017,Blais2021} and spin qubits in semiconductor quantum dots \cite{Loss1998}. Spin qubits are attractive due to their small size and the use of electron or hole spins as an intrinsic two-level system. Superconducting qubits, by contrast, benefit from efficient control and coupling via circuit quantum electrodynamics (circuit QED). Both platforms are regarded as scalable, as they are compatible with established semiconductor fabrication technologies.

A recent topic of interest is superconducting spin qubits \cite{Chtchelkatchev2003}, which combine the spin qubits and superconducting qubits into a single platform. These hybrid qubits make use of the Andreev bound states that arise in quantum dots in a Josephson junction, and are thus referred to as Andreev spin qubits. 
They can be coupled to transmon qubits and to each other using circuit QED \cite{Chtchelkatchev2003,Padurariu2010}, thereby addressing the difficult question of how to couple spin qubits across long distances. In this decade, there has been substantial experimental progress \cite{Hays2020,Hays2021, Hays2021Thesis, Bargerbos2023, PitaVidal2023, PitaVidal2024}, ranging from successful spin detection \cite{Hays2020} to strong coupling between two Andreev spin qubits \cite{PitaVidal2024}. A recent blueprint \cite{PitaVidal2025} proposed a setup that enables all-to-all connectivity between these qubits, building on earlier theoretical work on superconducting qubits \cite{Makhlin2001}. If realized, the enhanced connectivity would greatly simplify the implementation of error correction codes and quantum algorithms.

A two-quantum-dot variation of a spin qubit, the singlet-triplet qubit~\cite{Levy2002, Petta2005}, has the advantage of relying only on a single control mechanism: the exchange interaction between two quantum dots. However, singlet-triplet qubits still have the same limitation as conventional spin qubits. Since the two-qubit interaction is only short-range, the connectivity is limited.

In this paper, we propose and discuss superconducting singlet-triplet (SST) qubits. An SST qubit consists of two quantum dots that are placed in parallel in a single Josephson junction. This architecture shares several key advantages of Andreev spin qubits, such as long-range coupling, compatibility with transmon circuits, and all-to-all connectivity. Unlike Andreev spin qubits, however, SST qubits do not rely on spin–orbit interaction, thereby allowing the use of materials without intrinsic spin–orbit coupling, such as silicon.  These properties put them in line with other recent hybrid qubit proposals that use fermion parity \cite{Geier2024} or Yu-Shiba-Rusinov singlet states \cite{Steffensen2025} of a serial double quantum dot in a Josephson junction. 

The coupling between the SST qubit and the superconducting degrees of freedom naturally arises from the singlet-type of pairing of the superconductor. 
The coupling is based on crossed Andreev processes \cite{Choi2000, Choi2001, Scherubl2019, Spethmann2024}, where Cooper pairs split into different quantum dots, a mechanism demonstrated before in Cooper pair splitters \cite{Hofstetter2009, Baba2018, Ranni2021, Kurtossy2022, Recher2001}  and Andreev molecules \cite{Junger2023, Kurtossy2021}.  Relying on Cooper pair splitting processes may seem cumbersome; however, Andreev spin qubits likewise require Cooper pairs to split and traverse different dot orbitals \cite{Padurariu2010}. In our SST qubit, allowing the split Cooper pairs to propagate through separate dots enhances controllability.

Leijnse and Flensberg \cite{Leijnse2013} also proposed a singlet-triplet qubit based on crossed Andreev processes, but their qubit is not placed in a Josephson junction and is therefore independent of the superconducting phase or magnetic fluxes. In contrast, our SST qubit can be controlled by adjusting only magnetic fluxes, in analogy to the idea that conventional singlet-triplet qubits can be controlled  by adjusting only the exchange interaction.  

The number of flux lines required to control SST qubits scales unusually slowly. For an array of $N$ SST qubits, operating single-qubit gates (in parallel) and two-qubit gates between any two qubits (sequentially) requires control of only $N$ magnetic fluxes. In addition, $\mathcal{O}(N)$ voltage lines are needed for initialization in the computational basis, while readout can be performed using a single transmon qubit coupled to a microwave resonator. Here we focus on a readout protocol that requires all qubits to be measured. However, by analogy with Andreev spin qubits, more flexible readout schemes are expected to be feasible. Besides that, we also show that SST qubits are protected against both charge and flux noise to linear order.
Experimentally, the SST qubits could be realized in proximitized two-dimensional electron or hole gases, or with parallel double nanowires in a Josephson junction \cite{Junger2023, Kurtossy2021, Matsuo2022, Ueda2019, Kanne2022}.

\section{A singlet-triplet qubit in a double-dot Josephson junction} \label{sec:microscopics}

\begin{figure}
    \centering
    \includegraphics[width=0.5\linewidth]{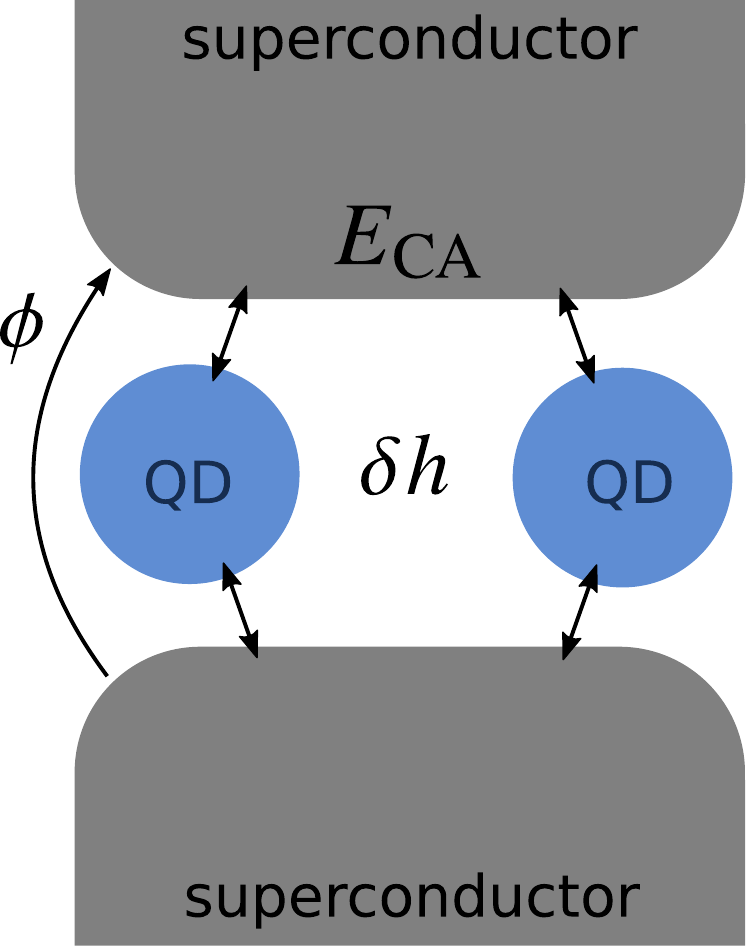}
    \caption{A superconducting singlet-triplet qubit consisting of a double quantum dot in a Josephson junction.  The spins in the quantum dots are coupled by crossed Andreev processes across the junction, $E_\text{CA}(1+\cos\phi)$, whose magnitude can be controlled by the superconducting phase $\phi$. Additionally, a difference in the dots' Zeeman energies $\delta h$ mixes singlet and triplet states. 
    }
    \label{fig:double_dot_josephson_junction}
\end{figure}

One SST qubit is made of a gate-defined double quantum dot in a Josephson junction, as shown in Fig. \ref{fig:double_dot_josephson_junction}. Both dots are tunnel coupled to the two superconducting leads with equal coupling constant $t$, but not to each other. The superconductors have a superconducting gap $\Delta$ and differ by a superconducting phase difference of $\phi$. The dots have one energy level with energy $\epsilon<0$ relative to the chemical potential and strong on-site Coulomb repulsion $\mathcal{U}$, favoring states where each dot is occupied by one electron. In addition, a magnetic field with a gradient splits those dot states by the Zeeman energies $h_L,h_R\lesssim |\epsilon|$, which differ by $\delta h=h_L-h_R$ between the dots.  We consider the parameters to fulfill $\Gamma, |\delta h| \ll |\epsilon|\ll \mathcal{U},\Delta$ where $\Gamma=\rho_F \pi t^2$ and $\rho_F$ is the normal-state density of states per spin of the leads at the Fermi level.
It was shown \cite{Choi2000, Choi2001, Scherubl2019, Spethmann2024}   
that under these assumptions the spins of the quantum dots interact with isotropic interaction 
\footnote{Choi et al. \cite{Choi2000} calculated the result in Eq.~\eqref{eq:H_DDJJ} for $\Delta\ll\mathcal{U}$ and without Zeeman field. However, their result stays valid without these restrictions, thus, for $\Gamma, h_{L/R}\ll |\epsilon|\ll\mathcal{U}$ and $\Gamma,h_{L/R}\ll \Delta$. The reason is that (i) a larger $\Delta$ does not enable new tunnel paths and (ii) Zeeman fields only appear in the energy denominators of perturbation theory, where they are much smaller than the excitation energies $\mathcal{O}(\Delta, |\epsilon|)$.},
\begin{align}\label{eq:H_DDJJ}
    &H_\text{DDJJ} \\\nonumber 
    &=E_{\text{CA}}\left[1+\cos(\phi)\right]\frac{1}{4}(\vec{\sigma}_L\cdot\vec{\sigma}_R-1) + \frac{h_L}{2}\sigma_L^z +\frac{h_R}{2}\sigma_R^z.
\end{align}
The interaction is proportional to $E_{\text{CA}}=2\Gamma^2/|\epsilon|$ and mediated by Cooper pairs. The Pauli vectors $\vec{\sigma}_L$ and $\vec{\sigma}_R$ describe the spins on the quantum dots, which are both occupied by one electron.
Since the interaction is isotropic, the triplet states $T_+=\ket{\uparrow\uparrow}$ and $T_-=\ket{\downarrow\downarrow}$ are eigenstates, decoupled from the rest. We define an SST qubit in the remaining subspace. In the basis $(\ket{\uparrow\downarrow},\ket{\downarrow\uparrow})$ with corresponding Pauli matrices $\tau^x$ and $\tau^z$, the Hamiltonian that describes one qubit is
\begin{align}\label{eq:single-qubit-Hamiltonian-standalone}
    &H'_\text{DDJJ}= \frac{E_{\text{CA}}}{2}[1+\cos(\phi)](\tau^x - 1) + \frac{\delta h}{2}\tau^z \\
    &= \!\left[\frac{E_{\text{CA}}}{2} \tau^x\!+\! \frac{\delta h}{2}\! \tau^z \right]- \frac{E_{\text{CA}}}{2}[1 + \cos(\phi)] + \frac{E_{\text{CA}}}{2}\cos(\phi)\tau^x. \nonumber
\end{align}
As with conventional singlet-triplet qubits, the Zeeman energy difference $\delta h$ splits the states in one direction (here: $z$ direction) and stays fixed. The spin interaction proportional to $E_{\text{CA}}$ enables qubit rotations around a different axis (here: $x$ axis). In conventional singlet-triplet qubits, the interaction arises from the exchange interaction and can be controlled by the detuning energy between the dots \cite{Jirovec2021} or the tunnel barrier between the dots. Here, the interaction is mediated by Cooper pairs as a result of crossed Andreev processes, and can be fully switched on or off with the superconducting phase difference $\phi$. The phase difference is controlled using magnetic fluxes that penetrate SQUID-like circuits, as explained in more detail in the next section. In the following, we will assume that the Zeeman energy difference is smaller than the crossed Andreev coupling coefficient, $\delta h \ll E_{\text{CA}}$, to drive single- and two-qubit gates.

\subsection{Initialization}

The ground state in the junction is the $T_-$ state, which is not in the computational basis. To initialize the SST qubit in the computational basis, we suggest increasing the crossed Andreev coupling $E_\text{CA}$ by raising the tunnel couplings $t$ between the quantum dots and the superconductors using the barrier gates. The requirement $\Gamma\ll|\epsilon|$ may be relaxed during this process, as we argue in App.~\ref{Sec:general_H_largeGamma}. When the superconducting phase difference is set to $\phi=0$ \footnote{Setting the superconducting phase difference to $\phi=0$ corresponds to the X flux set point described in Sec. \ref{sec:qubit_gates}.} and the crossed Andreev coupling is larger than the Zeeman energies, $E_{\text{CA}}\gtrsim h_{L,R}$, the singlet state becomes the ground state. Initialization is accomplished by waiting for the system to relax into that ground state. To operate the qubit, the singlet state is ramped back up to its original state by lowering the tunnel couplings. The crossed Andreev coupling needs to fulfill $ E_{\text{CA}}\lesssim h_{L},h_R$ to avoid leakage into the $T_-$ or $T_+$ state during the qubit operations.

\section{Qubit architecture controlled by fluxes}\label{sec:qubit_gates}

\begin{figure}
    \centering
    \includegraphics[width=0.98\linewidth]{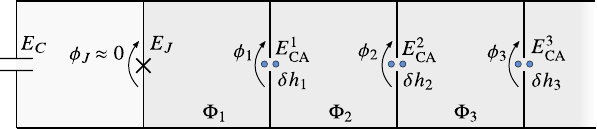}
    \caption{To operate $N$ SST qubits, they are placed in parallel in a superconducting circuit where they are controlled by magnetic fluxes $\Phi_j$, analogous to a setup suggested for Andreev spin qubits \cite{PitaVidal2025}. A conventional Josephson junction with large Josephson energy $E_J$ enables two-qubit interactions.
    }
    \label{fig:circuit_labels}
\end{figure}

A quantum computer requires many qubits. To realize both single- and two-qubit gates for each of them, the SST qubits can all be placed in parallel in a superconducting circuit, as shown in Fig.~\ref{fig:circuit_labels}. In addition, a conventional Josephson junction with large Josephson energy $E_J$ is connected in parallel, as well as a capacitor with capacitance $e^2/(2E_C)$ needed for readout \footnote{The $e$ is the elementary charge.}. 
The setup was originally proposed for the superconducting qubits~\cite{Makhlin2001} and adopted for Andreev spin qubits \cite{PitaVidal2025}. Below, we analyze its benefits for SST qubits. 

In the following, we call the setup an SST device. We number the qubits $j=1,..., N$ and describe them by the Pauli operators $\tau_j^x$ and $\tau_j^z$. Each of the qubits is controlled by the corresponding superconducting phase difference $\phi_j$, crossed Andreev coupling energy $E_\text{CA}^j$ and Zeeman energy difference $\delta h_j$, see Fig.~\ref{fig:circuit_labels}.
Magnetic fluxes $\Phi_j$ penetrate the circuit areas between the SST qubits $j-1$ and $j$ and can be controlled with nearby flux lines. They are related to the reduced flux $\varphi_j$ via $\varphi_j=\frac{2\pi}{\Phi_0}{\Phi_j}$, where $\Phi_0=\frac{\hbar \pi}{e}$ is the magnetic flux quantum \footnote{The $\hbar=\frac{h}{2\pi}$ is the reduced Planck's constant.}.
The Kirchhoff rules relate the phases to each other, $\phi_j = \phi_J+\varphi_j^s$, where $\phi_J$ is the superconducting phase across the conventional Josephson junction, and $\varphi_j^s=\sum_{l=1}^{j}\varphi_l$ is the sum of fluxes between the conventional junction and the qubit.
As a result, $\phi_J$ is the only remaining phase degree of freedom. We also define its conjugate charge number operator $n_q$.
The system Hamiltonian reads \footnote{We dropped the constant $-\frac{1}{2}\sum_{j=1}^NE_\text{CA}^j$.}
\begin{align}  \label{eq:initial-hamiltonian}
    H =\, & 4 E_C n_q^2 + E_J(1- \cos\phi_J)  + \sum_{j=1}^{N} \frac{E_\text{CA}^j}{2}(\tau_j^x -1)\cos\phi_j \notag \\
    &+ \frac{1}{2}\sum_{j=1}^N\left( E_\text{CA}^j\tau_j^x+\delta h_j \tau_j^z\right).
\end{align}

To analyze this Hamiltonian, we define the vector of Pauli-$x$ operators $\boldsymbol{\tau} = \{\tau_1^x,\dots,\tau_N^x\}$. Further, we denote the full qubit configuration by $\ket{\boldsymbol{s}} = \ket{s_1,\dots,s_N}$, where $s_j=\pm$ is the eigenvalue of the Pauli operator $\tau_j^x$ at site $j$, such that $\tau_j^x \ket{\boldsymbol{s}} = s_j \ket{\boldsymbol{s}}$. 

The $\phi_J$-dependent terms in Eq.~\eqref{eq:initial-hamiltonian} originate from the Josephson junctions in the circuit. Since all $\tau_j^x$ operators commute, these terms can be combined, leading to 
\begin{align}  \label{eq:Hamiltonian_combined_cosine}
    H =\, & 4 E_C n_q^2 + E_J - E_{J,\text{tot}}(\vec{\tau})\cos\left[\phi_J - \phi_{\text{tot}}(\vec{\tau})\right]  \notag \\
    &+ \frac{1}{2}\sum_{j=1}^N\left( E_\text{CA}^j\tau_j^x+\delta h_j \tau_j^z\right).
\end{align}
We see that the Josephson junctions together behave like a qubit-dependent SQUID 
with energy $E_{J,\text{tot}}(\vec{\tau})$ and offset $\phi_\text{tot}(\vec{\tau})$. The explicit dependence of $E_{J,\text{tot}}(\vec{\tau})$ and $\phi_\text{tot}(\vec{\tau})$ on the circuit parameters $E_\text{CA}^j$, $E_J$ and $\varphi_j^s$ is given in App. \ref{app:many-spin}. 

The circuit is operated in the transmon regime, where $\sqrt{8E_J E_C}\ll E_J$. The first line of Eq.~\eqref{eq:Hamiltonian_combined_cosine} describes the transmon dynamics. We assume that the qubit energy scales $\delta h_j$ and $E_{\text{CA}}^j$ are much smaller than the transmon frequency $\hbar\omega_{\text{tr}}(\vec{s})=\sqrt{8E_C E_{J,\text{tot}}(\vec{s})}\sim\sqrt{8 E_C E_J}$. As a result, excitations from the transmon ground state caused by the qubit dynamics are suppressed, and we can take a projection to that ground state. 
For each spin configuration $\ket{\vec{s}}$, the transmon ground state is localized around the minimum of the cosine-term, which is at $-E_{J,\text{tot}}(\vec{s})$. 
Thus, the projection to the ground state corresponds to expanding $-E_{J,\text{tot}}(\vec{\tau})$ (for $E_J\gg E_C$) up to second order. We find the qubit Hamiltonian
\begin{align} \label{eq:HSST}
H_\text{SST}=&\sum_{j=1}^{N}\sum_{k=j+1}^{N} H_{jk}^{2q} + \sum_{j=1}^N H_j^{1q} + H^{0q},\\\nonumber
H_{jk}^{2q}=&-\frac{E_\text{CA}^j E_\text{CA}^k}{4E_J}\sin (\varphi_j^s) \sin (\varphi_k^s) \, \tau_j^x\tau_k^x,\\\nonumber
H_{j}^{1q}=&\frac{E_\text{CA}^j}{2}\tau_j^x\left(1+\cos \varphi_j^s+\frac{1}{2} \sin \varphi_j^s\sum_{k=1}^N\frac{E_\text{CA}^k \sin \varphi_k^s}{E_J}\right) \\ \nonumber
&+ \frac{\delta h_j}{2} \tau_j^z.
\end{align}
Single- and two-qubit gates are enabled by $H^{1q}_j$ and $H^{2q}_{jk}$, respectively. The qubit-independent term $H^{0q}$ does not play a role here (see App. \ref{app:many-spin}). 
Formally, the projection corresponds to a Schrieffer-Wolff transformation where we treat the Zeeman term $\frac{1}{2}\sum_{j=1}^N \delta h_j \tau_j^z$ as a perturbation. Details of the calculation are described in App. \ref{app:many-spin}.

\subsection{Single-qubit gates}

\begin{figure*}
    \centering
    \includegraphics[width=0.98\linewidth]{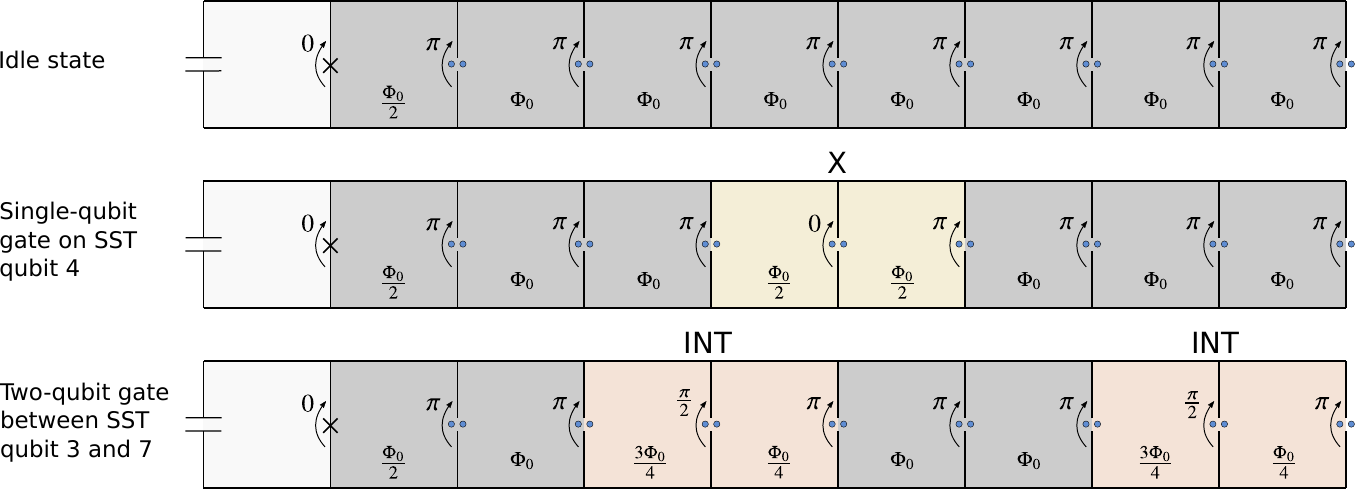}
    \caption{Superconducting singlet-triplet (SST) qubits are operated by changing the fluxes through the adjacent circuit loops, which change the superconducting phases across them (black arrows). Qubits in their OFF flux set points are in an idle state. Qubits in their X flux set point are subject to single-qubit rotations. Single-qubit rotations can be performed for different qubits at once. Any two qubits that are in their INT flux set point interact and can realize a two-qubit gate. }
    \label{fig:circuit_all_to_all_connected}
\end{figure*}

The superconducting phase across an SST qubit is described by the sum $\varphi_j^s$ of fluxes between the conventional junction and the qubit. 
When the fluxes are tuned to $\varphi_j^s=\pi$, we call it the OFF set point for qubit $j$. In this OFF set point, the one-qubit Hamiltonian becomes $H_{j}^{1q}=\frac{\delta h_j}{2} \tau_j^z$ and the qubit is in an idle state, only split in the $z$ direction by the Zeeman energy difference. 
If, instead, the sum of fluxes is $\varphi_j^s=0$, the Hamiltonian becomes $H_{j}^{1q}=\frac{E_\text{CA}^j}{2}\tau_j^x+\frac{\delta h_j}{2} \tau_j^z$. For 
$|\delta h_j|\ll E_\text{CA}^j$, this flux set point realizes an $x$ rotation and we call it the X set point for qubit $j$, see Fig.~\ref{fig:circuit_all_to_all_connected}.   The OFF set point and the X set point combined enable arbitrary qubit rotations \footnote{Moving into a rotated frame, given by the transformation $U_R(t)=e ^{-i\delta h\tau_j^zt/(2\hbar)}$, transforms $\tau_x$ in $H_j^{1q}$ into $\left[\cos(\delta h t/\hbar)\tau_j^x-\sin(\delta h t/\hbar)\tau_j^y\right]$. Therefore, a flux pulse applied at time $t=\frac{\pi \hbar N_t}{|\delta h|}$, $N_t \in \mathbb{N}$ enables an $x$ rotation, and a flux pulse at $t=\frac{\pi \hbar (N_t+1/2)}{|\delta h|}$ enables a $y$ rotation.}. 
Note that the flux set point for one qubit can be changed without affecting the flux set points of the other qubits by adjusting only the two adjacent fluxes. The flux $\Phi_j$ sets the target set point for qubit $j$, while tuning  $\Phi_{j+1}$ in the opposite direction retains the set points for the remaining qubits. 
Furthermore, the two-qubit interaction $H_{jk}^{2q}$ vanishes for both the OFF and the X set point. Therefore, two-qubit terms do not disturb single-qubit rotations, and single-qubit rotations can be performed in parallel.  

\subsection{Two-qubit gates} \label{sec:two-qubit-gates}

Two-qubit gates require an interaction between qubits $j$ and $k$. To turn that interaction on, the flux sums corresponding to the qubits $j$ and $k$ are both set to $\varphi_j^s=\varphi_k^s=\frac{\pi}{2}$, which we call the INT flux set points, while all other qubits remain in their OFF (or X) set point, see Fig.~\ref{fig:circuit_all_to_all_connected}. The interaction $H_{jk}^{2q}$ then becomes maximally strong and the two qubits are described by the Hamiltonian 
\begin{align}
H^\text{INT}_{jk}=&-\frac{E_\text{CA}^j E_\text{CA}^k}{4E_J}\tau_j^x\tau_k^x + \frac{\delta h_j}{2}\tau_j^z + \frac{\delta h_k{2}}\tau_k^z \\
&+  \frac{E_\text{CA}^j}{2}\left(1+\frac{\!E_\text{CA}^j\!+\!E_\text{CA}^k\!} {2E_J}\right)\tau_j^x \nonumber \\
&+ \frac{E_\text{CA}^k}{2}\left(1+\frac{\!E_\text{CA}^j\!+\!E_\text{CA}^k\!}{2E_J}\right)\tau_k^x . \nonumber
\end{align}
This Hamiltonian enables entangling gates. If $|\delta h_j|\ll \frac{E_\text{CA}^j E_\text{CA}^k}{E_J}$, we can obtain a controlled-Z gate (CZ) with an interaction pulse of length $T= \frac{\pi\hbar  E_J}{E_\text{CA}^j E_\text{CA}^k}$ (up to single-qubit rotations
\footnote{For example, we get a CZ gate with 
$CZ = e^{i\pi/4} H_j H_k R_k(j) R_j(k) e^{-i T H^\text{INT}_{jk} / \hbar} H_j H_k$. 
Here, $H_j$ is a Hadamard gate on qubit $j$ and $R_k(j)= \exp\left(i\frac{\pi}{4}\tau_j^x\left[\frac{2E_J+E_\text{CA}^{j}}{E_\text{CA}^{k}}\right]\right)$ is a rotation around the $x$ axis of qubit $j$.}). 
If $\delta h$ is not much smaller than $\frac{E_\text{CA}^j E_\text{CA}^k}{E_J}$, then one may need several interaction pulses for a CNOT or CZ gate,
and the exact implementation depends on the ratio between $\delta h_{j}$, $E_\text{CA}^j$ and $E_J$ \footnote{The calculation of Makhlin invariants \cite{Makhlin2002} may be helpful in finding the exact gate sequence.}.

\subsection{Noise resilience}

The idle OFF state is robust against flux perturbations and charge noise to linear order, but sensitive to hyperfine noise. 
To study the effect of flux noise, we expand the single-qubit Hamiltonian $H_j^{1q}$ [Eq.~\eqref{eq:HSST}] around the OFF set point at $\varphi_j^s=\pi$, and see directly that the first-order correction vanishes. Therefore, small fluctuations of the magnetic fluxes hardly impact the idle state. The effect of global flux noise can be further reduced by a twisted geometry of the loops \cite{PitaVidal2025}.

Charge noise may also be present in the device and cause variations in the dot energies $\epsilon$. These variations affect the magnitude of the crossed-Andreev coupling energy $E_\text{CA}$ but do not change the phase dependence of $H_j^{1q}$, and thus do not disturb the qubits' OFF states (see App.~\ref{Sec:general_H_largeGamma}).
Charge noise can also affect the tunnel couplings and cause the proximitized pairing amplitude $\Gamma$ of one superconductor to deviate by $\delta \Gamma$ from the pairing amplitude of the other superconductor. This deviation $\delta \Gamma$ appears as a flux-independent contribution $\propto \frac{\delta \Gamma^2}{\epsilon} \tau_j^x$ in the single-qubit Hamiltonian $H_{j}^{1q}$ (see App.~\ref{Sec:general_H_largeGamma}). Since the correction is quadratic in $\delta \Gamma$, the idle state is  protected to linear order from charge noise affecting the tunnel amplitudes.

We expect that the idle state will be primarily sensitive to dephasing caused by magnetic noise, such as hyperfine noise. Depending on the material, isotopic purification can reduce the effect of hyperfine noise.

\section{Readout}\label{sec:readout}

The transmon in the SST device can be used to read out the SST qubits using the standard dispersive readout of superconducting qubits. We note that the method we analyze here only allows one to read out all the qubits at the end of a quantum algorithm, since reading out one qubit adds uncontrolled phases to all other qubits. 
We expect that coupling the SST qubit to a fluxonium qubit, rather than to a transmon, would enable selective readout without disturbing the other SST qubits. Such a scheme was analyzed for Andreev spin qubits in Ref.~\cite{PitaVidal2025}, and the underlying mechanism is expected to carry over to the present case. Nevertheless, in this work we focus on a transmon-based device, as it leads to a simpler circuit architecture and provides a natural starting point for extending the analysis to more complex readout schemes.

The steps for dispersive readout are standard in circuit QED: The circuit is capacitively coupled to a microwave cavity~\cite{Blais2021,Krantz2019},
leading to a dispersive shift of the cavity frequency that depends on both the transmon and qubit states. The cavity frequency in turn can be probed by a waveguide, because photons propagating through the waveguide scatter off the cavity and acquire a scattering phase. Due to the dispersive shift, the scattering phase depends on the SST qubit states and can be measured afterwards, which completes the dispersive readout protocol. 

The capacitive coupling of the SST device to the cavity is described by the following Hamiltonian:
\begin{align}  \label{eq:ro-hamiltonian}
  H_{\text{RO}} = \hbar \omega_r b^{\dagger} b + \sum_{n=0}^{\infty} E_{\text{tr},n}(\vec{\tau})P_n + g (b + b^{\dagger})(a_{\vec{\tau}} + a^{\dagger}_{\vec{\tau}}).
\end{align}
Here, $\omega_r$ is the cavity frequency, $b^{\dagger}$ creates a cavity photon, $E_{\text{tr},n}(\vec{\tau})$ is the qubit-dependent transmon eigenvalue, $P_n $ is the projection operator to the $n$-th excitation of the transmon (see Eq.~\eqref{eq:projectors}), $a^{\dagger}_{\vec{\tau}}$  is the transmon creation operator, and $g$ is the capacitive coupling strength. 

In the dispersive regime, $|\hbar\omega_r - \hbar\omega_{\text{tr}}(\vec{s})| \gg g$, a Schrieffer-Wolff transformation can be used to perturbatively diagonalize the Hamiltonian $H_{\text{RO}}$~\eqref{eq:ro-hamiltonian}, resulting in an effective Hamiltonian of the cavity and the transmon \cite{Blais2021}:
\begin{align} \label{eq:dispersive-hamiltonian}
H_{\mathrm{disp}} =\ &
\hbar\omega_r\, b^\dagger b \\
&
+\sum_{n\ge 0}\left[\hbar\omega_{\text{tr}}(\boldsymbol s)\, n + \Lambda_n(\boldsymbol s) + \chi_n(\boldsymbol s)\,b^\dagger b\right] P_n. \notag
\end{align}
Here, $\Lambda_n(\boldsymbol{s})/\hbar$ is the Lamb shift of the transmon frequency, and $\chi_n(\boldsymbol{s})/\hbar$ is the dispersive shift of the cavity frequency. The detailed expressions of $\Lambda_n(\boldsymbol{s})$ and $\chi_n(\boldsymbol{s})$ are given in App.~\ref{app:sst-cavity}.
The key term in the Hamiltonian in Eq.~\eqref{eq:dispersive-hamiltonian} is the dispersive shift $\chi_n(\boldsymbol{s})$. Via this term, the spin states affect the frequency of the cavity. The SST qubits are operated in the transmon ground state, in which case the dispersive shift is
\begin{align}  \label{eq:effective-cavity-frequency}
    \chi_0(\boldsymbol{s})
    \approx  - \frac{g^2}{\Xi_0} - \frac{g^2}{(\Xi_0)^2}\sqrt{\frac{E_C}{2E_J}}\sum_{j=1}^N E_\text{CA}^j(\tau_j^x - 1)\cos\varphi_j^s,
\end{align}
where $\Xi_0 = \sqrt{8E_CE_J}-E_C-\hbar\omega_r$. The dispersive shift influences the scattering phase of the waveguide photons, described by the $S$ matrix
\begin{equation} \label{eq:s-matrix}
    S_{12}(\omega_p) = \frac{\kappa/2}{\kappa/2 - i(\hbar\omega_p - [\hbar\omega_r + \chi_0(\boldsymbol{s})])}.
\end{equation}
Here, $\kappa$ is the rate at which the cavity photons decay into the waveguide, and $\omega_p$ is the frequency of the probe photon. When the frequency of the waveguide photons is close to the cavity frequency $\omega_p \approx \omega_r$, the phase of the scattered photons depends on the qubit states and can be measured. 

\begin{figure}
    \centering
\hspace*{0.5cm}\includegraphics[width=0.9\linewidth]{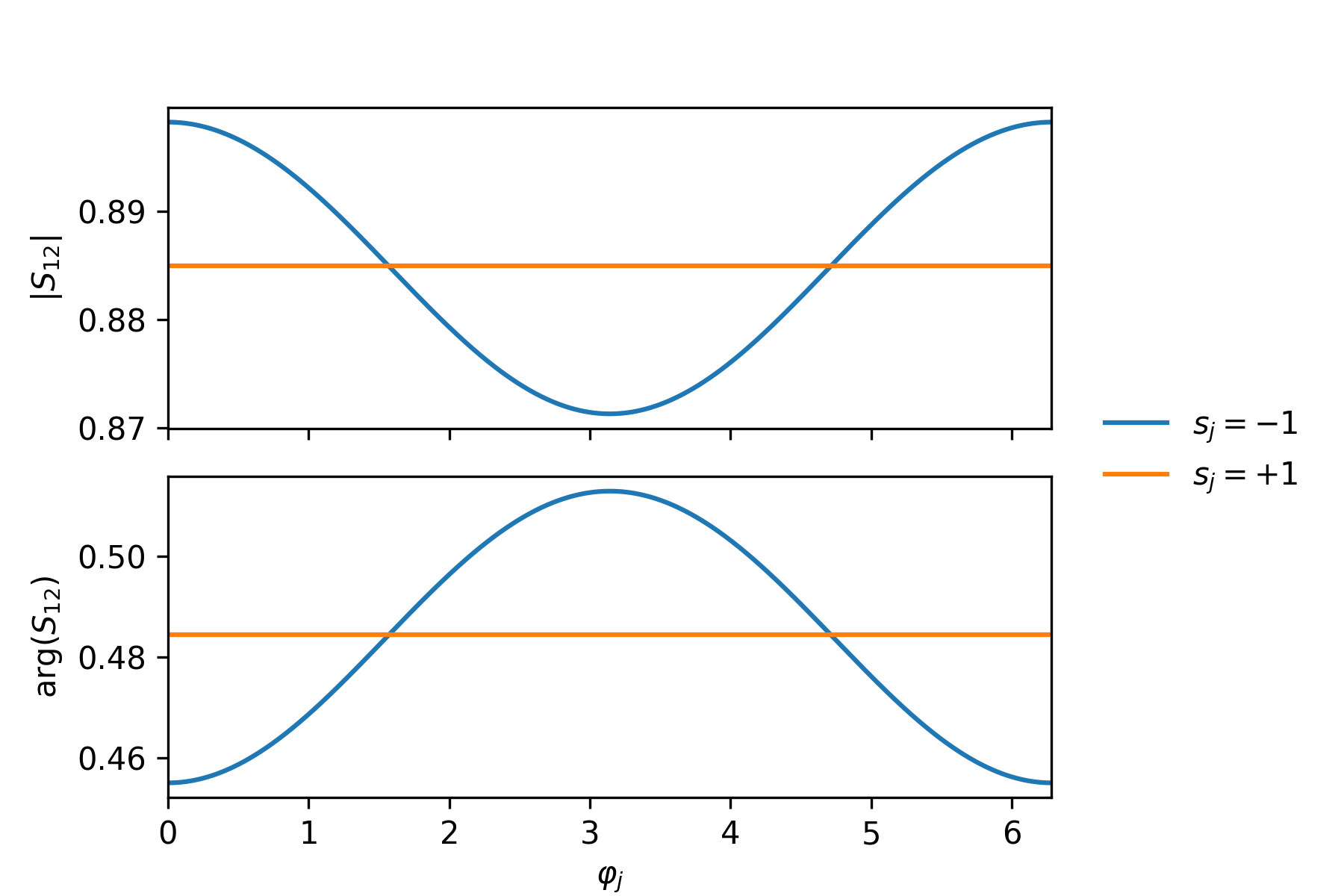}
    \caption{The $S_{12}$ coefficient [Eq.~\eqref{eq:s-matrix}] of the scattered waveguide photons as a function of the magnetic flux $\varphi_j^s$. The $S_{12}$ coefficient is plotted for two possible states of the $j$th qubit $s_j=\pm1$. The other magnetic fluxes are tuned to the INT set point ($\varphi_{k\neq j}^s = \pi/2$) to make the respective qubits not affect the photon scattering phase. The plot clearly shows that the maximum difference between the scattering phases for the qubit states $s_j = \pm 1$ is achieved in the OFF set point ($\varphi_j^s = \pi$) and the X set point ($\varphi_j^s = 0$). The $S_{12}$ coefficient is plotted for the following parameter values: $\kappa/h = 20\, \mathrm{MHz}$, $\omega_p/2\pi =\omega_r/2\pi = 4\, \mathrm{GHz}$, $E_J/h = 45\, \mathrm{GHz}$, $E_C/h = 100\, \mathrm{MHz}$, $E_{\text{CA}}/h = 2\, \mathrm{GHz}$, $g/h = 100\, \mathrm{MHz}$.}
    \label{fig:s-matrix}
\end{figure}

Equation~\eqref{eq:effective-cavity-frequency} shows that the sensitivity of the cavity frequency to a given SST qubit is controlled by the sum of fluxes $\varphi_j^s$, similarly to the single- and two-qubit gates. SST qubits in their OFF or X set points ($\varphi_j^s = \pi$ or $\varphi_j^s = 0$) contribute to the cavity frequency shift and are therefore measurable, while qubits at the INT set point ($\varphi_j^s = \pi/2$) are effectively invisible to the cavity, see Fig.~\ref{fig:s-matrix}.

To distinguish the measurement outcomes of different qubits, the following approach can be pursued, in analogy to Ref.~\cite{PitaVidal2025}. All qubits are tuned to their INT set point. Then, one qubit at a time is temporarily set to its OFF set point and its measurement outcome is noted. Setting all qubits to their INT set point at the beginning will let them all interact and acquire difficult-to-control phases. However, since they are measured in the same basis ($x$ basis) afterwards, these controlled $x$-rotations do not affect their measurement outcome. 

The requirement that all the qubits must be read out can be an obstacle to error correction codes that require repeated measurement of ancilla qubits. Selective readout of individual SST qubits could be achieved by placing the qubits in a fluxonium circuit and using higher-order transitions as in Ref.~\cite{PitaVidal2025}. Selective readout could also be achieved by using the barrier gates to decouple all the singlet-triplet qubits that are not being measured from the superconducting circuit \footnote{In that case, the Hamiltonian describing one qubit simply becomes $H_\text{DDJJ}'=\frac{\delta h}{2}\tau_j^z$ (for $\Gamma=0$), see App.~\ref{Sec:general_H_largeGamma}.}.

The dispersive readout also puts an additional constraint on the Zeeman energy difference of the qubits. Since the readout basis is not an eigenbasis of the Hamiltonian (due to the Zeeman term), the qubits will perform a $z$-rotation during the readout. Thus, the described readout protocol requires $\delta h\ll h/T_\text{read}$, where $T_\text{read}$ is the required readout time, restricting the single-qubit gate speed. We expect a similar tradeoff to also exist for Andreev spin qubits.

Nevertheless, the simple and controllable way of coupling the SST qubits to a transmon highlights their compatibility with the circuit QED architecture. Our results therefore establish a solid foundation for the development of more flexible and selective readout schemes in the future.

\section{Conclusion}

In this work we propose a hybrid semiconductor-superconductor qubit architecture, the superconducting singlet-triplet (SST) qubit. For the semiconductor singlet-triplet qubit, the control is implemented by a tunable exchange coupling. The proposed SST qubit is a conventional singlet-triplet qubit placed in a Josephson junction, which leads to the exchange term being coupled to the supercurrent. Therefore, the SST qubit is controlled through superconducting degrees of freedom, which allows us to leverage the established superconducting-qubit toolbox for SST operation.

Compared to the Andreev spin qubit architecture~\cite{Chtchelkatchev2003}, the SST qubit design does not rely on spin-orbit interaction and instead uses an inhomogeneous magnetic field [cf. Sec.~\ref{sec:microscopics}]. This makes the approach compatible with hybrid devices based on silicon. The drawback is that two quantum dots are required per SST qubit compared to a single quantum dot per Andreev spin qubit, complicating the device fabrication. On the other hand, the SST operation involves only the lowest orbital level in each dot (whereas the Andreev spin qubit relies on two orbital levels), which relaxes the requirements on the individual quantum dot quality.

A further potential advantage is tunable all-to-all connectivity in an SST qubit array [cf. Sec.~\ref{sec:qubit_gates}]. For multiple SST qubits embedded in a common transmon, two-qubit gates between arbitrary pairs are mediated by the shared transmon mode. The addressed qubit pair is selected by tuning the corresponding external flux biases.

Finally, we investigated a readout scheme that relies on dispersive readout of the coupled transmon qubit [cf. Sec.~\ref{sec:readout}]. Because dispersive readout outcomes are not aligned with the SST eigenbasis, the readout is destructive. Therefore, the readout speed constrains the admissible strength of the SST Zeeman term, which in turn limits single-qubit gate speeds. On the other hand, a similar readout-related tradeoff also arises for Andreev spin qubits.

Despite these challenges, the SST qubit provides a promising route to hybrid semiconductor-superconductor devices that circumvents several limitations of Andreev spin qubits.

\section{Acknowledgements} This work was supported as part of NCCR SPIN, a National Center of Competence in Research, funded by the Swiss National Science Foundation (grant number 225153). D.L.~acknowledges the Deanship of Research and the Quantum Center at KFUPM for the support received under Grant no.~CUP25102 and no.~INQC2600, respectively.

\appendix

\section{Qubit-phase coupling based on crossed Andreev processes }\label{Sec:general_H_largeGamma}

Here, we review and combine the calculations from Refs.\cite{Choi2000, Choi2001, Scherubl2019, Spethmann2024} that lead to Eq.~\eqref{eq:single-qubit-Hamiltonian-standalone}. We clarify the conditions under which the perturbation theory is valid, justifying that the condition $\Gamma\ll|\epsilon|$ can be relaxed for initialization.
We also discuss the results for unequal dot energies and tunnel couplings, which are needed to understand the effects of charge noise. 

We start with the Hamiltonian of two superconductors tunnel coupled to two quantum dots. The dots are described by the Hamiltonian
\begin{align}
H_\text{QD}=\sum_{\alpha\sigma}\left(\epsilon_{\alpha}+\sigma\frac{h_{\alpha}}{2}\right)d_{\alpha\sigma}^{\dagger}d_{\alpha\sigma} + \mathcal{U}\sum_{\alpha} d_{\alpha\uparrow}^{\dagger}d_{\alpha\uparrow}d_{\alpha\downarrow}^{\dagger}d_{\alpha\downarrow}.
\end{align}
Here, $d_{\alpha\sigma}^{\dagger}$ creates an electron on dot $\alpha\in\{L,R\}$ with spin $\sigma\in\{\uparrow,\downarrow\}$ (or $\sigma\in\{+1,-1\}$),  negative dot energy $\epsilon_{\alpha}<0$ (measured from the chemical potential) and Zeeman energy $0<h_{\alpha}<|2\epsilon_{\alpha}|$. There is an on-site Coulomb repulsion $\mathcal{U}$.

The Hamiltonian of the superconductors is the mean-field BCS Hamiltonian
\begin{align}
H_\text{SC}=\sum_{n\vec{k}\sigma}\xi_kc_{n\vec{k}\sigma}^{\dagger}c_{n\vec{k}\sigma}-\sum_{n\vec{k}}\Delta e^{-i\phi^{\text{sc}}_n}c_{n\vec{k}\uparrow}^\dagger c_{n\vec{-k}\downarrow}^{\dagger} + \text{H.c.}
\end{align}
Here, $c^{\dagger}_{n\vec{k}\sigma}$ creates an electron on superconductor $n\in\{1,2\}$ with momentum $\vec{k}$, spin $\sigma$, and dispersion relation $\xi_k$. The superconducting gap $\Delta$ describes the pairing potential, along with the superconducting phases $\phi^{\text{sc}}_{1/2}=\pm\phi/2$. The Hamiltonian can be diagonalized with a Bogoliubov transformation $c_{n\vec{k}\sigma}=u_k\gamma_{n\vec{k}\sigma}+\sigma v_{nk}\gamma^{\dagger}_{n-\vec{k}-\sigma}$ with amplitudes $u_k=\sqrt{\frac{E_k+\xi_k}{2E_k}}$, $v_{nk}=e^{-i\phi^{\text{sc}}_n}\sqrt{\frac{E_k-\xi_k}{2E_k}}$, and energy $E_k=\sqrt{\Delta^2 +\xi_k^2}$:
\begin{align}
    H_\text{SC}=\sum_{n\vec{k}\sigma}E_k\gamma_{n\vec{k}\sigma}^{\dagger}\gamma_{n\vec{k}\sigma} + \text{const.}
\end{align}
The superconductors and quantum dots are tunnel coupled with real, spin-conserving tunnel coupling constants $t_{n\alpha}>0$:
\begin{align}
    H_\text{T}=\sum_{n\alpha\sigma\vec{k}}t_{n\alpha}c^{\dagger}_{n\vec{k}\sigma}d_{\alpha\sigma} + \text{H.c.}
\end{align}

\begin{figure}
\includegraphics[width=0.9\linewidth]{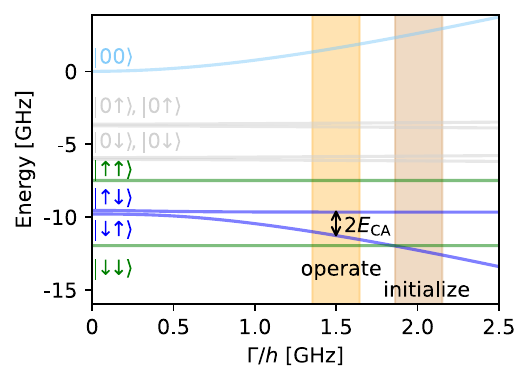}
    \caption{Spectrum of the states in a Josephson junction at a superconducting phase of $\phi=0$. 
    The computational states (dark blue) are split by the Crossed Andreev coupling energy $2E_\text{CA}=4\frac{\Gamma^2}{|\epsilon|}$, because the computational singlet state couples to the empty-dot state (light blue) through crossed Andreev processes [see Eq.~\eqref{eq:H_even_2}]. The polarized triplet states $\ket{\uparrow\uparrow}$ and $\ket{\downarrow\downarrow}$ (green) remain uncoupled, just like the one-electron states $\ket{0\sigma}$ and $\ket{\sigma0}$ (gray). When $E_\text{CA}\gtrsim h_\alpha/2$, a singlet state becomes the ground state, which can be used to initialize the qubit (brown-shaded region). For qubit operations (in the orange-shaded region) it is important that $\ket{\downarrow\downarrow}$ is outside of the qubit energy window to avoid leakage. We chose $\epsilon_L/h=\epsilon_R/h=-4.8\,$GHz, $\Gamma_1=\Gamma_2=\Gamma$, $h_L/h=-2.4\,$GHz, $h_R/h=2.2\,$GHz, $\Delta/h=48\,$GHz.
    }
    \label{fig:spectrum_initialization}
\end{figure}

Now, we perform second-order perturbation theory \cite{Winkler2003} for large Coulomb repulsion and superconducting gap: $\Gamma_{n}, \Gamma_{n\alpha}^\text{LA},h_{\alpha}, \epsilon_{\alpha} \ll \mathcal{U}, \Delta$. We defined $\Gamma_{n}=t_{nL} t_{nR}\pi\rho_F$, $\Gamma_{n\alpha}^{\text{LA}}=t_{n\alpha}^2\pi\rho_F$, and the density of states per spin at the Fermi level  $\rho_F$ 
\footnote{We assume that the density of states in the superconductors is constant, $\rho(\epsilon)=\rho_F$, and in some calculations we take the principal value of the integral.}. 
All states with no more than one electron per quantum dot will be well separated from states with at least one doubly occupied dot, or states with an excitation at a superconductor. The Hamiltonian conserves parity and spin. 
Therefore, the two singly occupied states $\ket{0\sigma}$ and $\ket{\sigma0}$ only couple to each other
\footnote{The coupling energy is $\frac{-\Gamma_1-\Gamma_2}{4\Delta}(\sum_{\alpha}\epsilon_{\alpha}+\sigma\frac{h_{\alpha}}{2})$.} 
and the two triplet states $T_+=\ket{\uparrow\uparrow}$ and $T_-=\ket{\downarrow\downarrow}$ remain uncoupled with energies $\sum_{\alpha}\epsilon_\alpha\pm  h_\alpha/2$ 
\footnote{The superconductors renormalize the dot energies $\epsilon_{L,R}$ and Zeeman energies $h_{L,R}$ of the states, so from now on, the energies $\epsilon_\alpha$ and $h_{\alpha}$ refer to the corresponding normalized energies. Here, we assumed that the density of states in the leads is symmetric around the Fermi energy.}. The remaining three even parity states are described by the Hamiltonian [in the basis $(\ket{00},\ket{\uparrow\downarrow},\ket{\downarrow\uparrow})$]:
\begin{align}\label{eq:H_even_2}
H_{even}^{(2)}=\begin{pmatrix} -\Sigma\epsilon &\Gamma_\phi^*&-\Gamma_\phi^*\\
\Gamma_\phi&(h_L-h_R)/2&0\\
-\Gamma_\phi&0 &(-h_L+h_R)/2\end{pmatrix} + \Sigma\epsilon. 
\end{align}
Here we defined the sum of the dot energies, $\Sigma\epsilon=\epsilon_L+\epsilon_R$, and the proximitized pairing $\Gamma_\phi=-\sum_n \Gamma_ne^{-i\phi^{\text{sc}}_n}$. 
 Next, we perform another second-order perturbation theory into the states $\ket{\uparrow\downarrow}$ and $\ket{\downarrow\uparrow}$, assuming $|\Gamma_\phi|\ll |\Sigma\epsilon\pm\frac{\delta h}{2}|$. We arrive at a more general version of Eq.~\eqref{eq:single-qubit-Hamiltonian-standalone}:
\begin{align}\label{eq:HDDJJprime_long}
H_\text{DDJJ}'=&
\left( \frac{\delta h}{2}-|\Gamma_\phi|^2 \frac{\delta h/2}{(\Sigma\epsilon)^2-(\delta h/2)^2}\right)\tau^z\nonumber
\\
&+ |\Gamma_\phi|^2 \frac{\Sigma\epsilon}{(\Sigma\epsilon)^2-(\delta h/2)^2} (1-\tau^x).
\end{align}
We obtain the solution from the main text for $\Gamma_L=\Gamma_R$, $\delta h\ll |\epsilon_\alpha|$ and $\epsilon_L=\epsilon_R$.

Eq.~\eqref{eq:HDDJJprime_long} shows what happens when the parameters deviate from these values, and how sensitive the results are with respect to noise. Unequal dot energies $\epsilon_L\neq \epsilon_R$ do not qualitatively affect the result, as only the sum $\Sigma\epsilon$ enters Eq.~\eqref{eq:HDDJJprime_long}. For $\delta h \cancel{\ll}|\epsilon_\alpha|$, the x-rotation at the X set point will be tilted towards the $z$ direction, and similarly, the two-qubit interaction will contain $\tau_x^{(j)}\tau_z^{(k)}$-terms and $\tau_z^{(j)}\tau_z^{(k)}$-terms. Finding the right pulse sequence to realize a certain single- or two-qubit gate may then require numerical tools. 
To analyze $\Gamma_1\neq\Gamma_2$, it is helpful to define $\bar{\Gamma}=(\Gamma_1+\Gamma_2)/2$ and $\delta\Gamma=(\Gamma_1-\Gamma_2)/2$, resulting in 
\begin{equation}
    |\Gamma_{\phi}|^2=2(\bar{\Gamma}^2-\delta \Gamma^2)(1+\cos\phi) +4\delta \Gamma^2. 
\end{equation}
Even though noise that causes $\delta\Gamma\neq0$ disturbs the idle state of the qubit (at $\phi=\pi$) because the coefficient of $(1-\tau_x)$ does not vanish anymore, the correction is quadratic in $\delta\Gamma$ and is therefore suppressed for small $\delta \Gamma$.

The Hamiltonian $H_\text{even}^{(2)}$ of the three even parity states [Eq.~\eqref{eq:H_even_2}] is suited to describe the qubit initialization in the computational space. For large $|\Gamma_\phi|\sim |\epsilon_\alpha|,h_\alpha$ the hybridization of $\ket{\uparrow\downarrow}$ and $\ket{\downarrow\uparrow}$ with $\ket{00}$ dominates,  and one of the states replaces $T_-$ as the ground state, see Fig~\ref{fig:spectrum_initialization}. Quickly lowering $|\Gamma_\phi|$ again will turn that strongly hybridized state into $\ket{\uparrow\downarrow}$ or $\ket{\downarrow\uparrow}$ during a diabatic transition, which concludes the initialization procedure.

\section{Many-qubit system} \label{app:many-spin}

\subsection{A qubit-dependent SQUID} 

A SQUID-array of SST qubits connected in parallel with a transmon qubit enables single- and two-qubit gates, as argued in Section \ref{sec:qubit_gates}. In this Appendix we describe in detail how to obtain the qubit Hamiltonian $H_{\text{SST}}$ in Eq.~\eqref{eq:HSST} from the full system Hamiltonian in Eq.~\eqref{eq:initial-hamiltonian}, a calculation involving a Schrieffer-Wolff transformation.
In the following, we will call the SST qubit states `spin' states, since they are two-level systems. They should not be confused with the electron spins in the quantum dots.

The system Hamiltonian $H_{\text{SST}}$ in Eq.~\eqref{eq:initial-hamiltonian} describes the circuit SST device in Fig.~\ref{fig:circuit_labels}. To analyze it, we first consider only the terms involving the superconducting phase $\phi_J$. Together they form an effective SQUID. While these terms explicitly depend on the spin states, the picture of a SQUID gives an intuitive understanding of its behavior. The resulting expression reads
\begin{align}\label{eq:spin-squid}
        &V_{\text{SQUID}}(\phi_J, \boldsymbol{\tau}) \\
        &\equiv E_J(1- \cos(\phi_J))+ \sum_{j=1}^N\frac{E_\text{CA}^j}{2}\cos(\phi_j)(\tau_j^x -1) \nonumber \\
        &=E_J -E_{J,\text{tot}}(\boldsymbol{\tau})\cos(\phi_J - \phi_{\text{tot}}(\boldsymbol{\tau})). \nonumber
\end{align}
Here, we have introduced the total Josephson energy $E_{J,\text{tot}}(\vec{\tau})$ and the total superconducting phase offset $\phi_{\text{tot}}(\vec{\tau})$. Since these quantities depend on the operators $\tau_j^x$ (which commute for different sites $j$), the energy $E_{J,\text{tot}}(\vec{\tau})$ and the phase offset $\phi_{\text{tot}}(\vec{\tau})$ are diagonal in the spin basis $\{\ket{\boldsymbol{s}}\}$. Explicitly, they read 
\begin{widetext}
\begin{align} \label{eq:spin-dependent-josephson-energy-and-phase}
    E_{J,\text{tot}}(\boldsymbol{\tau}) &= \sqrt{\left[E_J - \sum_{j=1}^N\frac{E_\text{CA}^j}{2} (\tau_j^x-1)\cos(\varphi_j^s)\right]^2 +   \left[\sum_{j=1}^N\frac{E_\text{CA}^j}{2}(\tau_j^x-1)\sin(\varphi_j^s)\right]^2} \\
    \phi_{\text{tot}}(\boldsymbol{\tau}) &= \arctan\left[\frac{\sum_{j=1}^N \frac{E_\text{CA}^j}{2}(\tau_j^x - 1)\sin(\varphi_j^s)}{E_J - \sum_{j=1}^N \frac{E_\text{CA}^j}{2} (\tau_j^x-1) \cos(\varphi_j^s)}\right].
\end{align}
\end{widetext}
Via the introduced notation, the full Hamiltonian of the system reads
\begin{equation}
\label{eq:full-system-hamiltonian}
\begin{aligned}
    &H = 4 E_C n_q^2 + E_{J,\text{tot}}(\boldsymbol{\tau})[1-\cos(\phi_J - \phi_{\text{tot}}(\boldsymbol{\tau}))] \\
    &+ [E_J - E_{J,\text{tot}}(\boldsymbol{\tau})] +  \sum_{j=1}^N \frac{E_\text{CA}^j}{2}(\tau_j^x - 1) + \sum_{j=1}^N \frac{\delta h_j}{2}\tau_j^z.
\end{aligned}
\end{equation}
This expression separates into a transmon part (first line) and a spin part (second line). The spin part contains the term $[E_J - E_{J,\text{tot}}(\vec{\tau})]$, which is the spin-dependent energy minimum of the combined SQUID, 
\begin{equation}
    \min_{\phi_J}V_{\text{SQUID}}(\phi_J) = E_J - E_{J,\text{tot}}(\boldsymbol{\tau}).
\end{equation}
This term plays a key role in the SST device since it is responsible for the two-qubit gates discussed in Sec.~\ref{sec:two-qubit-gates}. For $E_J \gg \sum_j E_\text{CA}^j$, it can be expanded as 
\begin{align}\label{eq:expand_E_J_tot}
E_J -E_{J,\text{tot}}(\vec{\tau}) \approx & \sum_{j=1}^N\frac{E_\text{CA}^j}{2}(\tau_j^x - 1)\cos(\varphi_j^s)  \\
& - \frac{1}{8E_J}\left[\sum_{j=1}^N E_\text{CA}^j (\tau_j^x - 1)\sin(\varphi_j^s)\right]^2. \nonumber
\end{align}
This term provides the leading two-qubit interaction, as we will argue in the remaining part of this appendix. However, the validity of this expansion depends not only on the crossed Andreev reflection energy $E_\text{CA}$ but also on the number of qubits. This restricts the number of SST qubits that can be placed in a single array.

\subsection{Transmon regime}

The transmon part of the Hamiltonian is given by 
\begin{align}\label{eq:transmon_part_of_Hamiltonian}
    H_{\text{tr}} &= 4 E_C n_q^2 + E_{J,\text{tot}}(\boldsymbol{\tau})[1-\cos(\phi_J - \phi_{\text{tot}}(\boldsymbol{\tau}))],
\end{align}
and we analyze it for each spin configuration $|\vec{s}\rangle$ separately.
We define the spin-dependent creation and annihilation operators
\begin{align}\label{eq:creation_and_annihilation_operators}
  a_{\boldsymbol{\tau}} &= \frac{1}{2}\sqrt[4]{\frac{E_{J,\text{tot}}(\boldsymbol{\tau})}{2E_C}} \left[\phi_J -\phi_{tot}(\boldsymbol{\tau})\right]+ i\sqrt[4]{\frac{2E_C}{E_{J,\text{tot}}(\boldsymbol{\tau})}}n_q, \nonumber\\
  a_{\boldsymbol{\tau}}^{\dagger} &= \frac{1}{2}\sqrt[4]{\frac{E_{J,\text{tot}}(\boldsymbol{\tau})}{2E_C}} \left[\phi_J -\phi_{tot}(\boldsymbol{\tau})\right] - i\sqrt[4]{\frac{2E_C}{E_{J,\text{tot}}(\boldsymbol{\tau})}}n_q.
\end{align}
Assuming the transmon regime $E_C \ll E_J$, we can approximate $H_\text{tr}$ as an anharmonic oscillator. This requires expanding the cosine around its minimum and neglecting unbalanced terms with an unequal number of creation and annihilation operators, leading to %
\cite{Blais2021}\footnote{We dropped a spin-independent constant of $-\frac{E_C}{4}$}
\begin{align}\label{eq:transmon-hamiltonian}
    H_{\text{tr}} \approx \hbar\omega_{\text{tr}}(\boldsymbol{\tau}) \left(a_{\boldsymbol{\tau}}^{\dagger} a_{\boldsymbol{\tau}} + \frac{1}{2}\right)  - E_C a_{\vec{\tau}}^\dagger a_{\vec{\tau}}- \frac{E_C}{2} a_{\boldsymbol{\tau}}^{\dagger} a_{\boldsymbol{\tau}}^{\dagger} a_{\boldsymbol{\tau}} a_{\boldsymbol{\tau}}.
\end{align}
Since for different spin states $\ket{\vec{s}}$ we obtain different boson creation and annihilation operators $a^{\dagger}_{\vec{s}}$ and $a_{\vec{s}}$, the boson occupation number states $\ket{n_{\vec{s}}}$ also depend on the spins, 
\begin{equation}
    a^{\dagger}_{\boldsymbol{s}} a_{\boldsymbol{s}}\ket{n_{\boldsymbol{s}}} = n \ket{n_{\boldsymbol{s}}}.
\end{equation}
Thus, for each spin state $\ket{\vec{s}}$, the transmon energies are given by
\begin{align} \label{eq:transmon_energies}
E_{\text{tr},n}(\vec{s})=\hbar \omega_{\text{tr}}(\boldsymbol{s})\left(n + \frac{1}{2}\right) - \frac{E_C}{2}n(n+1).
\end{align}

We combine the transmon part of the Hamiltonian with the longitudinal spin part into
\begin{align}
    H_0 = H_{\text{tr}} + [E_{J,\text{tot}}(\boldsymbol{\tau}) - E_J] + \sum_{j=1}^N\frac{E_\text{CA}^j}{2} (\tau_j^x-1),
\end{align}
which is the full Hamiltonian without the Zeeman term.
This Hamiltonian has the eigenstates
\begin{equation}
    \ket{n, \boldsymbol{s}} \equiv \ket{n_{\boldsymbol{s}}} \otimes \ket{\boldsymbol{s}} 
\end{equation}
and can be directly written in its diagonal form
$H_0 =\sum_{n,\vec{s}} E_{n\vec{s}} |n,\vec{s}\rangle \langle n,\vec{s}|$ with eigenvalues
\begin{align} 
E_{n, \boldsymbol{s}} = E_{\text{tr},n}(\vec{s}) + [E_{J,\text{tot}}(\boldsymbol{s}) - E_J] + \sum_{j=1}^N \frac{E_\text{CA}^j}{2}s_j.
\end{align}
The spin configuration $\vec{s}$ and the occupation number $n$ appear as quantum numbers in the eigenvalues.

\subsection{Perturbation theory - Part I}\label{app:SW2}

The Zeeman term
\begin{equation}
  V = \sum_{j=1}^N \frac{\delta h_j}{2}\tau_j^z
\end{equation}
is not diagonal in $\{|n,\vec{s}\rangle\}$ and we treat $V$ as a small perturbation, assuming $\delta h_j\ll \hbar \omega_{\text{tr}}(\vec{s})$. Thus, the full Hamiltonian is a sum of the unperturbed Hamiltonian with known eigenvalues and eigenvectors, and the perturbation,
\begin{align}
    H = H_0 + V.
\end{align}
We employ a Schrieffer--Wolff (SW) transformation to decouple states with different boson numbers, yielding an effective Hamiltonian that is block-diagonal in $n$. To this end, we introduce projectors onto subspaces with fixed boson number:
\begin{equation}\label{eq:projectors}
  P_n = \sum_{\boldsymbol{s}\in\{+, -\}^{N}} \ket{n,\boldsymbol{s}}\bra{n,\boldsymbol{s}}.
\end{equation}
The effective Hamiltonian is obtained by a unitary transformation $H_{\text{eff}} = e^{S_{\text{SW}}} H e^{-S_{\text{SW}}}$, where $S_{\text{SW}}$ is an anti-Hermitian operator such that $H_\text{eff}$ becomes block-diagonal, $H_\text{eff}=\sum_n P_n H_\text{eff} P_n$.

We decompose the perturbation into block-diagonal (bd) and block-off-diagonal (bod) parts relative to the subspaces:
\begin{equation}
  V_{\text{bd}} = \sum_{n} P_n V P_n, \qquad
  V_{\text{bod}} = \sum_{n\neq m} P_n V P_m .
\end{equation}
The standard SW condition to eliminate the first-order block-off-diagonal terms is:
\begin{equation}
  [H_0,S_{\text{SW}}] = - V_{\text{bod}}.
\end{equation}
This condition determines the matrix elements of the generator $S_{\text{SW}}$ in the eigenbasis of $H_0$:
\begin{equation}
  \bra{n',\vec{s}'} S_{\text{SW}} \ket{n,\vec{s}}
  = - \frac{\bra{n',\vec{s}'} V_{\text{bod}} \ket{n,\vec{s}}}
         {E_{n',\vec{s}'} - E_{n,\vec{s}}},
  \qquad n'\neq n.
\end{equation}
The resulting effective Hamiltonian is given by
\begin{equation}\label{eq:schrieffer_wolff_expression_second_order}
  H_{\text{eff}} = H_0 + V_{\text{bd}} + \frac{1}{2}[S_{\text{SW}}, V_{\text{bod}}]
  + \mathcal{O}(\delta h^3).
\end{equation}
We see that the first order term is entirely given by $V_\text{bd}$, which is of the order of $\mathcal{O}(\delta h)$, whereas the second-order term scales at most with $\mathcal{O}(\frac{\delta h^2}{\hbar \omega_\text{tr}})$. 

At this point, it is helpful to determine what the matrix elements of $V_\text{bd}$ and $V_\text{bod}$ are. The reason why the perturbation has block-off-diagonal elements is that after a spin-$z$ operator has acted on a spin state, the spin state does not match the spin-dependent occupation number state any more:
\begin{align}\label{eq:bare_qubit_operator_behavior}
\tau_j^z\ket{n,\vec{s}}=\tau_j^z(\ket{n_{\vec{s}}}\ket{\vec{s}})=\ket{n_{\vec{s}}}\ket{\vec{s}'}\neq\ket{n_{\vec{s}'}}\ket{\vec{s}'}=\ket{n,\vec{s}'}. 
\end{align}
Here we defined $\ket{\vec{s'}}$ to be the same spin state as $\ket{\vec{s}}$ except at the spin component $j$, where $s'_j=-s_j$. 
We see that evaluating the matrix elements of $V_\text{bd}$ and $V_\text{bod}$ involves calculating the state overlap $\braket{n'_{\vec{s}'}|n_{\vec{s}}}$. We find the state overlap to be
\begin{align} \label{eq:number_bases_overlap_scaling_long}
    \braket{n'_{\vec{s}'}|n_{\vec{s}}}&\approx \begin{cases} 1 - \mathcal{O}(u^2 ,v^2 ), &\text{ for } n=n'\\ \mathcal{O}(u,v)  &\text{ for } n\neq n'\end{cases}
\end{align}
where $u$ and $v$ are small and dimensionless.
We devote the following subsection to proving Eq.~\eqref{eq:number_bases_overlap_scaling_long} and providing the scaling for $u$ and $v$. We continue with the Schrieffer-Wolff calculation in App. \ref{app:SW_continued}.

\subsection{Number state overlap of different qubit states}

To evaluate the Schrieffer-Wolff transformation, we need to calculate the overlap $\braket{n'_{\vec{s}'}|n_{\vec{s}}}$ and prove Eq.~\eqref{eq:number_bases_overlap_scaling_long}. As a reminder, here $\vec{s}$ and $\vec{s}'$ are the same except for $s'_j=-s_j$ where $j$ is fixed.
For this purpose, we only study the harmonic oscillator part of the transmon Hamiltonian (Eq.~\eqref{eq:transmon_part_of_Hamiltonian})
\begin{align}
H_\text{ho} &= 4 E_C n_q^2 +\frac{1}{2} E_{J,\text{tot}}(\vec{\tau})[\phi_J- \phi_\text{tot}(\vec{\tau})]^2 \nonumber \\
&=  \hbar\omega_{\text{tr}}(\boldsymbol{\tau}) \left(a_{\boldsymbol{\tau}}^{\dagger} a_{\boldsymbol{\tau}} + \frac{1}{2}\right).
\end{align}
Since the creation and annihilation operators are defined in the same way as for the transmon Hamiltonian (Eq.~\eqref{eq:creation_and_annihilation_operators}), the boson occupation number states $\ket{n_{\vec{s}}}$ are the same for the harmonic and for the anharmonic oscillator. Since we assume $E_J\gg E_C$, we remain in the low occupation sector and can neglect the $2\pi$-periodicity of the phase operator.

The creation and annihilation operators $\{a_{\vec{s}'},a_{\vec{s}'}^\dagger\}$ can be written as a function of $\{a_{\vec{s}},a_{\vec{s}}^\dagger\}$ via the phase operator $\phi_J$ and charge operator $n_q$. Using the definitions in Eq.~\eqref{eq:creation_and_annihilation_operators}, we know $a_{\vec{s}'}^{(\dagger)}=a_{\vec{s}'}^{(\dagger)}(\phi_J, n_q)$  and can write the phase and charge as $\phi_J=\phi_J(a_{\vec{s}},a_{\vec{s}}^\dagger)$ and $n_q=n_q(a_{\vec{s}},a_{\vec{s}}^\dagger)$.
To simplify the expressions, we define the squeezing operator $S(u)$ and the displacement operator $D(v)$, standard in quantum optics: 
\begin{align}
    \label{eq:squeezing-operator-and-displacement-operator}
  S(u) &= \exp\bigl[u \bigl( (a_{\vec{s}}^{\dagger})^2 - a_{\vec{s}}^2\bigr)\bigr],\\
  u &= \frac{1}{4}\ln\!\left(\frac{\omega_{\text{tr}}(\vec{s}')}{\omega_{\text{tr}}(\vec{s})}\right), \nonumber\\
    D(v) &= \exp\bigl[ v (a_{\vec{s}}^{\dagger} - a_{\vec{s}}) \bigr],\\
    v& = \frac{1}{4} \sqrt{\frac{\hbar}{E_C}}\left(\sqrt{\omega_{\text{tr}}(\vec{s}')} \phi_\text{tot}(\vec{s}') - \sqrt{\omega_{\text{tr}}(\vec{s})} \phi_\text{tot}(\vec{s})\right).\nonumber
\end{align}
The displacement operator $D(v)$ is unitary and changes the harmonic-oscillator displacement from $\phi_{\text{tot}}(\vec{s})$ to $\phi_\text{tot}(\vec{s}')$, because $D(v)a^{(\dagger)}_{\vec{s}}D(v)^{\dagger}= a^{(\dagger)}-v$ \cite{walls_quantum_2025}. The squeezing operator is also unitary and transforms the creation and annihilation operators according to $S(u)a_{\vec{s}}S^{\dagger}(u)=a_{\vec{s}} \cosh(2u) -a_{\vec{s}}^\dagger\sinh(2u)$ \cite{walls_quantum_2025}. Thus, it effectively changes the boson frequency from $\omega_\text{tr}(\vec{s})$ to $\omega_\text{tr}(\vec{s}')$. Combining these properties, we find:
\begin{align}  \label{eq:a-plus-from-a-minus}
  a_{\vec{s}'} &= S(u)\, D(v)\, a_{\vec{s}}\, D^\dagger(v)\, S^\dagger(u),\\
  a_{\vec{s}'}^{\dagger} &= S(u)\, D(v)\, a_{\vec{s}}^{\dagger}\, D^\dagger(v)\, S^\dagger(u).
\end{align}
Accordingly, the number bases are related by
\begin{equation} \label{eq:number-bases-2}
  \ket{n_{\vec{s}'}} = S(u) D(v)\ket{n_{\vec{s}}}.
\end{equation}

Assuming $E_\text{CA}^k\ll\sqrt{8E_C E_J}\ll E_J$, we can use the expansion of $E_{J,\text{tot}}(\vec{s})$ in Eq.~\eqref{eq:expand_E_J_tot} and
\begin{align}
    \phi_{\text{tot}}(\boldsymbol{\tau})\approx \sum_{k=1}^N \frac{E_\text{CA}^{k}}{2E_J}(\tau_k^x - 1) \sin(\varphi_k^s),
\end{align}
and show that the parameters $u$ and $v$ are small, scaling with
\begin{align}
u &\sim \frac{1}{8}\ln\left(\frac{E_J+\mathcal{O}( E_\text{CA}^j)}{E_J-\mathcal{O}({E_\text{CA}^j})}\right)\sim \mathcal{O}\left(\frac{E_\text{CA}^j}{E_J}\right),\\
v &\sim \mathcal{O}\left(\sqrt[4]{\frac{E_J}{E_C}}\times\frac{E_\text{CA}^j}{E_J}\right)\sim\mathcal{O}\left(\sqrt[4]{\frac{E_C}{E_J}}\times \frac{E_\text{CA}^j}{\hbar\omega_{\text{tr}}}\right),\nonumber
\end{align}
where $j$ is the number of the spin whose orientation differs for the spin configurations $\boldsymbol{s}$ and $\boldsymbol{s}'$.
As a result, we can expand the squeezing and displacement operators for $u,v\ll 1$ and approximate the overlap of the number states of different spin configurations
\begin{align}
&\braket{n_{\vec{s}}|n'_{\vec{s}'}}=\bra{n_{\vec{s}}}S(u)D(v)\ket{n_{\vec{s}}}\\ \nonumber
&\approx\!\bra{n_{\vec{s}}}[1+u(a_{\vec{s}}^{\dagger2} -a_{\vec{s}}^2 )] [1+v(a_{\vec{s}}^\dagger - a_{\vec{s}})]+\mathcal{O}(u^2, v^2)\ket{n_{\vec{s}}}.
\end{align}
Finally, we get the result  Eq.~\eqref{eq:number_bases_overlap_scaling_long}.

\subsection{Perturbation theory - Part II}\label{app:SW_continued}

Knowing the scaling of the occupation number state overlap, we can return to our calculation of the Schrieffer-Wolff transformation. Our goal was to calculate the block-diagonal $V_\text{bd}$ and block-off-diagonal $V_\text{bod}$ components of the Zeeman term, to evaluate Eq.~\eqref{eq:schrieffer_wolff_expression_second_order}. Since $V_\text{bod}$ scales with $\mathcal{O}(u,v)$, the second order term in Eq.~\eqref{eq:schrieffer_wolff_expression_second_order} is suppressed even more. It scales with $\mathcal{O}(\delta h^2 u^2/(\hbar\omega_\text{tr})^2)$ or $\mathcal{O}(\delta h^2 v^2/(\hbar\omega_\text{tr})^2)$,  giving us a good justification to neglect it completely. Further, the first-order term $V_\text{bd}$ is approximated to couple the number states with equal $n$ but different spin configurations perfectly. 

The effective Hamiltonian then becomes for any low-$n$ occupation sector:
\begin{align}  \label{eq:many-spin-effective-hamiltonian}
  H_{\text{eff}} \simeq& \hbar\omega_{\text{tr}}(\boldsymbol{\tau}) \left(a^{\dagger}_{\vec{\tau}} a_{\vec{\tau}} + \frac{1}{2}\right) - \frac{E_C}{2}a^{\dagger}_{\vec{\tau}} a_{\vec{\tau}} - \frac{E_C}{2} a^{\dagger}_{\vec{\tau}} a^{\dagger}_{\vec{\tau}} a_{\vec{\tau}} a_{\vec{\tau}} \notag \\
  &+ [E_{J,\text{tot}}(\boldsymbol{\tau}) - E_J] + \sum_{j=1}^N \frac{E_\text{CA}^j}{2} \tau_j^x + \sum_{j=1}^N \frac{\delta h_j}{2}\tau_j'^z.
\end{align}
This Hamiltonian looks almost identical to the unperturbed Hamiltonian. However, it differs in two ways. First, all operators now act on a transformed basis, $\ket{n,s}_\text{SW}=e^{S_\text{SW}}\ket{n,s}$. Second, we introduced a new qubit operator, $\tau_j^{\prime z}$, that, unlike $\tau_j^z$ in Eq.~\eqref{eq:bare_qubit_operator_behavior}, does not contain any transmon-number leakage terms by fulfilling $\tau_j^{\prime z}\ket{n,\vec{s}}=\ket{n,\vec{s}'}_\text{SW}$. As before, the $\vec{s}'$ only differs from $\vec{s}$ by $s_j=-s'_j$.

Finally, we specialize to the transmon ground state by projecting onto the $n=0$ sector using $H_{\text{spin}} =P_0 H_{\text{eff}}P_0$,
\begin{align}
  H_{\text{spin}} =&  \frac{1}{2}\hbar\omega_{\text{tr}}(\vec{\tau}) + [E_{J,\text{tot}}(\vec{\tau}) - E_J] \\
  &+ \frac{1}{2}\sum_{j=1}^N \left[E_\text{CA}^j \tau_j^x + \delta h_j \tau_j'^{z}\right].\nonumber
\end{align}

Two-qubit interaction terms, originating from $E_{J,\text{tot}}(\vec{\tau})$, are present both in the SQUID energy minimum $[E_{J,\text{tot}}(\vec{\tau})-E_J]$ and in the vacuum energy $\frac{1}{2}\hbar\omega_{\text{tr}}(\vec{\tau})$.
However, the vacuum energy is suppressed by a factor of $\sqrt{E_C/E_J}$ compared to the SQUID energy minimum. Therefore, we neglect it. Even if we did not neglect it, it would only renormalize the coefficients of the interaction, which in any case are obtained experimentally. 

Without the vacuum energy and using the expansion of $E_{J,\text{tot}}(\vec{\tau})$ (Eq.~\eqref{eq:expand_E_J_tot}), we obtain
\begin{align}
 H_\text{SST} \approx&\sum_{j=1}^N\frac{E_\text{CA}^j}{2}(\tau_j^x - 1)\cos(\varphi_j^s) \\\nonumber
 &- \frac{1}{2E_J}\left[\sum_{j=1}^N\frac{E_\text{CA}^j}{2}(\tau_j^x - 1)\sin(\varphi_j^s)\right]^2 \\ \nonumber
 &+ \frac{1}{2}\sum_{j=1}^N \left[E_\text{CA}^j \tau_j^x + \delta h_j \tau_j^{\prime z}\right].
\end{align}
This is the result in the main text (Eq.~\eqref{eq:HSST}), where we dropped the primes from the Pauli operators,  because we do not explicitly distinguish the different boson number states there. The spin-independent term in Eq.~\eqref{eq:HSST} is 
\begin{align}
H^{0q}= &-\sum_{j=1}^N\frac{E_\text{CA}^j}{2}\cos\varphi_j^s \nonumber\\
&-\frac{1}{8E_J}\sum_{j=1}^N\left(E_\text{CA}^j\right)^2\sin^2\varphi_j^s \nonumber\\
&-\frac{1}{8E_J}\left(\sum_{j=1}^N E_\text{CA}^{j}\sin\varphi_j^s\right)^2.
\end{align}
The Schrieffer-Wolff result coincides with the result obtained by minimizing the potential energy of the combined SQUID term in Eq.~\eqref{eq:spin-squid}. 
This agreement is not surprising, because the transmon ground state is localized at its potential minimum, but convenient because the capacitive terms of the system Hamiltonian (Eq.~\eqref{eq:initial-hamiltonian}) can simply be neglected. Thus, treating the superconducting phase $\phi_J$ as a classical variable whose energy needs to be minimized coincides with the full quantum treatment involving the Schrieffer-Wolff transformation.

\section{Coupling to a cavity} \label{app:sst-cavity}

In the main text we describe how dispersive readout of the SST qubits is accomplished by coupling the SST qubit device to a microwave  cavity. In this appendix we provide the expressions of the Lamb shift and the dispersive shift, which result from a Schrieffer-Wolff transformation. 

The starting Hamiltonian contains the effective SST Hamiltonian~\eqref{eq:many-spin-effective-hamiltonian} capacitively coupled to a cavity,
\begin{equation} \label{eq:cavity-sst}
    H_\text{disp} = H_{\text{eff}} + \hbar \omega_r b^{\dagger} b + g (a + a^{\dagger})(b + b^{\dagger}),
\end{equation}
where $\omega_r$ is the cavity frequency, $b^{\dagger}$ and $b$ are the cavity creation and annihilation operators, and $g$ is the capacitive coupling strength. 

For the dispersive regime, we assume that $|\omega_r - \omega_{\text{tr}}| \gg g$.
We first apply a rotating wave approximation (RWA) to eliminate the counter-rotating terms. Next, we use a Schrieffer-Wolff transformation to diagonalize the Hamiltonian~\eqref{eq:cavity-sst}  following Blais et al.~\cite{Blais2021}. We obtain the Hamiltonian in Eq.~\eqref{eq:dispersive-hamiltonian}, where, besides using the definition of the transmon energies $E_{\text{tr},n}(\vec{s})$ [Eq.~\eqref{eq:transmon_energies}] and the projectors to the $n$-occupation subspace $P_n$ [Eq.~\eqref{eq:projectors}], we identify the Lamb shift $\Lambda_n(\boldsymbol{s})$ and the dispersive shift $\chi_n(\boldsymbol{s})$ as 
\begin{align} \label{eq:dispersive-sst-parameters}
    \Lambda_n(\boldsymbol s) &\approx 
\frac{n g^2}{\Xi_{n-1}}\Bigg[1
+\sqrt{\frac{E_C}{2E_J}}\frac{\sum_{j=1}^N E_\text{CA}^j(\tau_j^x-1)\cos\varphi_j^s}{\Xi_{n-1}}
\Bigg], \notag\\
\chi_n(\boldsymbol s) &\approx \frac{n g^2}{\Xi_{n-1}}\Bigg[1
+\sqrt{\frac{E_C}{2E_J}}\frac{\sum_{j=1}^N E_\text{CA}^j(\tau_j^x-1)\cos\varphi_j^s}{\Xi_{n-1}} \Bigg]\notag \\
&\mkern-25mu -\frac{(n+1)g^2}{\Xi_{n}}\Bigg[1 
+ \sqrt{\frac{E_C}{2E_J}}\frac{\sum_{j=1}^N E_\text{CA}^j(\tau_j^x-1)\cos\varphi_j^s}{\Xi_{n}}
\Bigg].
\end{align}
We added the following shorthand for brevity:
\begin{equation} \label{eq:dispersive-sst-parameters-2}
\Xi_n =\sqrt{8E_CE_J}-(n+1)E_C-\hbar\omega_r .
\end{equation}
The dispersive shift $\chi_n(\vec{s})$ is particularly relevant because it shows how the flux set points affect the effective cavity frequency and enable the readout of the SST qubits.

\bibliography{literature}

\begin{thebibliography}{51}%
\makeatletter
\providecommand \@ifxundefined [1]{%
 \@ifx{#1\undefined}
}%
\providecommand \@ifnum [1]{%
 \ifnum #1\expandafter \@firstoftwo
 \else \expandafter \@secondoftwo
 \fi
}%
\providecommand \@ifx [1]{%
 \ifx #1\expandafter \@firstoftwo
 \else \expandafter \@secondoftwo
 \fi
}%
\providecommand \natexlab [1]{#1}%
\providecommand \enquote  [1]{``#1''}%
\providecommand \bibnamefont  [1]{#1}%
\providecommand \bibfnamefont [1]{#1}%
\providecommand \citenamefont [1]{#1}%
\providecommand \href@noop [0]{\@secondoftwo}%
\providecommand \href [0]{\begingroup \@sanitize@url \@href}%
\providecommand \@href[1]{\@@startlink{#1}\@@href}%
\providecommand \@@href[1]{\endgroup#1\@@endlink}%
\providecommand \@sanitize@url [0]{\catcode `\\12\catcode `\$12\catcode `\&12\catcode `\#12\catcode `\^12\catcode `\_12\catcode `\%12\relax}%
\providecommand \@@startlink[1]{}%
\providecommand \@@endlink[0]{}%
\providecommand \url  [0]{\begingroup\@sanitize@url \@url }%
\providecommand \@url [1]{\endgroup\@href {#1}{\urlprefix }}%
\providecommand \urlprefix  [0]{URL }%
\providecommand \Eprint [0]{\href }%
\providecommand \doibase [0]{https://doi.org/}%
\providecommand \selectlanguage [0]{\@gobble}%
\providecommand \bibinfo  [0]{\@secondoftwo}%
\providecommand \bibfield  [0]{\@secondoftwo}%
\providecommand \translation [1]{[#1]}%
\providecommand \BibitemOpen [0]{}%
\providecommand \bibitemStop [0]{}%
\providecommand \bibitemNoStop [0]{.\EOS\space}%
\providecommand \EOS [0]{\spacefactor3000\relax}%
\providecommand \BibitemShut  [1]{\csname bibitem#1\endcsname}%
\let\auto@bib@innerbib\@empty
\bibitem [{\citenamefont {Devoret}(1995)}]{Devoret1995}%
  \BibitemOpen
  \bibfield  {author} {\bibinfo {author} {\bibfnamefont {M.~H.}\ \bibnamefont {Devoret}},\ }\bibfield  {title} {\bibinfo {title} {Quantum fluctuations in electrical circuits},\ }\href {https://boulderschool.yale.edu/sites/default/files/files/devoret_quantum_fluct_les_houches.pdf} {\bibfield  {journal} {\bibinfo  {journal} {Les Houches, Session LXIII}\ } (\bibinfo {year} {1995})}\BibitemShut {NoStop}%
\bibitem [{\citenamefont {Vool}\ and\ \citenamefont {Devoret}(2017)}]{Vool2017}%
  \BibitemOpen
  \bibfield  {author} {\bibinfo {author} {\bibfnamefont {U.}~\bibnamefont {Vool}}\ and\ \bibinfo {author} {\bibfnamefont {M.}~\bibnamefont {Devoret}},\ }\bibfield  {title} {\bibinfo {title} {Introduction to quantum electromagnetic circuits},\ }\href {https://doi.org/https://doi.org/10.1002/cta.2359} {\bibfield  {journal} {\bibinfo  {journal} {Int. J. Circuit Theory Appl.}\ }\textbf {\bibinfo {volume} {45}},\ \bibinfo {pages} {897} (\bibinfo {year} {2017})}\BibitemShut {NoStop}%
\bibitem [{\citenamefont {Blais}\ \emph {et~al.}(2021)\citenamefont {Blais}, \citenamefont {Grimsmo}, \citenamefont {Girvin},\ and\ \citenamefont {Wallraff}}]{Blais2021}%
  \BibitemOpen
  \bibfield  {author} {\bibinfo {author} {\bibfnamefont {A.}~\bibnamefont {Blais}}, \bibinfo {author} {\bibfnamefont {A.~L.}\ \bibnamefont {Grimsmo}}, \bibinfo {author} {\bibfnamefont {S.~M.}\ \bibnamefont {Girvin}},\ and\ \bibinfo {author} {\bibfnamefont {A.}~\bibnamefont {Wallraff}},\ }\bibfield  {title} {\bibinfo {title} {Circuit quantum electrodynamics},\ }\href {https://doi.org/10.1103/RevModPhys.93.025005} {\bibfield  {journal} {\bibinfo  {journal} {Rev. Mod. Phys.}\ }\textbf {\bibinfo {volume} {93}},\ \bibinfo {pages} {025005} (\bibinfo {year} {2021})}\BibitemShut {NoStop}%
\bibitem [{\citenamefont {Loss}\ and\ \citenamefont {DiVincenzo}(1998)}]{Loss1998}%
  \BibitemOpen
  \bibfield  {author} {\bibinfo {author} {\bibfnamefont {D.}~\bibnamefont {Loss}}\ and\ \bibinfo {author} {\bibfnamefont {D.~P.}\ \bibnamefont {DiVincenzo}},\ }\bibfield  {title} {\bibinfo {title} {Quantum computation with quantum dots},\ }\href {https://doi.org/10.1103/PhysRevA.57.120} {\bibfield  {journal} {\bibinfo  {journal} {Phys. Rev. A}\ }\textbf {\bibinfo {volume} {57}},\ \bibinfo {pages} {120} (\bibinfo {year} {1998})}\BibitemShut {NoStop}%
\bibitem [{\citenamefont {Chtchelkatchev}\ and\ \citenamefont {Nazarov}(2003)}]{Chtchelkatchev2003}%
  \BibitemOpen
  \bibfield  {author} {\bibinfo {author} {\bibfnamefont {N.~M.}\ \bibnamefont {Chtchelkatchev}}\ and\ \bibinfo {author} {\bibfnamefont {Y.~V.}\ \bibnamefont {Nazarov}},\ }\bibfield  {title} {\bibinfo {title} {Andreev quantum dots for spin manipulation},\ }\href {https://doi.org/10.1103/PhysRevLett.90.226806} {\bibfield  {journal} {\bibinfo  {journal} {Phys. Rev. Lett.}\ }\textbf {\bibinfo {volume} {90}},\ \bibinfo {pages} {226806} (\bibinfo {year} {2003})}\BibitemShut {NoStop}%
\bibitem [{\citenamefont {Padurariu}\ and\ \citenamefont {Nazarov}(2010)}]{Padurariu2010}%
  \BibitemOpen
  \bibfield  {author} {\bibinfo {author} {\bibfnamefont {C.}~\bibnamefont {Padurariu}}\ and\ \bibinfo {author} {\bibfnamefont {Y.~V.}\ \bibnamefont {Nazarov}},\ }\bibfield  {title} {\bibinfo {title} {Theoretical proposal for superconducting spin qubits},\ }\href {https://doi.org/10.1103/PhysRevB.81.144519} {\bibfield  {journal} {\bibinfo  {journal} {Phys. Rev. B}\ }\textbf {\bibinfo {volume} {81}},\ \bibinfo {pages} {144519} (\bibinfo {year} {2010})}\BibitemShut {NoStop}%
\bibitem [{\citenamefont {Hays}\ \emph {et~al.}(2020)\citenamefont {Hays}, \citenamefont {Fatemi}, \citenamefont {Serniak}, \citenamefont {Bouman}, \citenamefont {Diamond}, \citenamefont {de~Lange}, \citenamefont {Krogstrup}, \citenamefont {Nyg{\aa}rd}, \citenamefont {Geresdi},\ and\ \citenamefont {Devoret}}]{Hays2020}%
  \BibitemOpen
  \bibfield  {author} {\bibinfo {author} {\bibfnamefont {M.}~\bibnamefont {Hays}}, \bibinfo {author} {\bibfnamefont {V.}~\bibnamefont {Fatemi}}, \bibinfo {author} {\bibfnamefont {K.}~\bibnamefont {Serniak}}, \bibinfo {author} {\bibfnamefont {D.}~\bibnamefont {Bouman}}, \bibinfo {author} {\bibfnamefont {S.}~\bibnamefont {Diamond}}, \bibinfo {author} {\bibfnamefont {G.}~\bibnamefont {de~Lange}}, \bibinfo {author} {\bibfnamefont {P.}~\bibnamefont {Krogstrup}}, \bibinfo {author} {\bibfnamefont {J.}~\bibnamefont {Nyg{\aa}rd}}, \bibinfo {author} {\bibfnamefont {A.}~\bibnamefont {Geresdi}},\ and\ \bibinfo {author} {\bibfnamefont {M.}~\bibnamefont {Devoret}},\ }\bibfield  {title} {\bibinfo {title} {Continuous monitoring of a trapped superconducting spin},\ }\href {https://doi.org/https://doi.org/10.1038/s41567-020-0952-3} {\bibfield  {journal} {\bibinfo  {journal} {Nature Physics}\ }\textbf {\bibinfo {volume} {16}},\ \bibinfo {pages} {1103} (\bibinfo {year} {2020})}\BibitemShut {NoStop}%
\bibitem [{\citenamefont {Hays}\ \emph {et~al.}(2021)\citenamefont {Hays}, \citenamefont {Fatemi}, \citenamefont {Bouman}, \citenamefont {Cerrillo}, \citenamefont {Diamond}, \citenamefont {Serniak}, \citenamefont {Connolly}, \citenamefont {Krogstrup}, \citenamefont {Nyg{\aa}rd}, \citenamefont {Yeyati}, \citenamefont {Geresdi},\ and\ \citenamefont {Devoret}}]{Hays2021}%
  \BibitemOpen
  \bibfield  {author} {\bibinfo {author} {\bibfnamefont {M.}~\bibnamefont {Hays}}, \bibinfo {author} {\bibfnamefont {V.}~\bibnamefont {Fatemi}}, \bibinfo {author} {\bibfnamefont {D.}~\bibnamefont {Bouman}}, \bibinfo {author} {\bibfnamefont {J.}~\bibnamefont {Cerrillo}}, \bibinfo {author} {\bibfnamefont {S.}~\bibnamefont {Diamond}}, \bibinfo {author} {\bibfnamefont {K.}~\bibnamefont {Serniak}}, \bibinfo {author} {\bibfnamefont {T.}~\bibnamefont {Connolly}}, \bibinfo {author} {\bibfnamefont {P.}~\bibnamefont {Krogstrup}}, \bibinfo {author} {\bibfnamefont {J.}~\bibnamefont {Nyg{\aa}rd}}, \bibinfo {author} {\bibfnamefont {A.~L.}\ \bibnamefont {Yeyati}}, \bibinfo {author} {\bibfnamefont {A.}~\bibnamefont {Geresdi}},\ and\ \bibinfo {author} {\bibfnamefont {M.~H.}\ \bibnamefont {Devoret}},\ }\bibfield  {title} {\bibinfo {title} {Coherent manipulation of an {Andreev} spin qubit},\ }\href {https://doi.org/10.1126/science.abf0345} {\bibfield  {journal} {\bibinfo  {journal} {Science}\ }\textbf {\bibinfo {volume}
  {373}},\ \bibinfo {pages} {430} (\bibinfo {year} {2021})}\BibitemShut {NoStop}%
\bibitem [{\citenamefont {Hays}(2021)}]{Hays2021Thesis}%
  \BibitemOpen
  \bibfield  {author} {\bibinfo {author} {\bibfnamefont {M.}~\bibnamefont {Hays}},\ }\href {https://doi.org/10.1007/978-3-030-83879-9} {\emph {\bibinfo {title} {Realizing an {Andreev} Spin Qubit: Exploring Sub-gap Structure in Josephson Nanowires Using Circuit {QED}}}},\ Springer Theses\ (\bibinfo  {publisher} {Springer},\ \bibinfo {address} {Cham},\ \bibinfo {year} {2021})\BibitemShut {NoStop}%
\bibitem [{\citenamefont {Bargerbos}\ \emph {et~al.}(2023)\citenamefont {Bargerbos}, \citenamefont {Pita-Vidal}, \citenamefont {\ifmmode~\check{Z}\else \v{Z}\fi{}itko}, \citenamefont {Splitthoff}, \citenamefont {Gr\"unhaupt}, \citenamefont {Wesdorp}, \citenamefont {Liu}, \citenamefont {Kouwenhoven}, \citenamefont {Aguado}, \citenamefont {Andersen}, \citenamefont {Kou},\ and\ \citenamefont {van Heck}}]{Bargerbos2023}%
  \BibitemOpen
  \bibfield  {author} {\bibinfo {author} {\bibfnamefont {A.}~\bibnamefont {Bargerbos}}, \bibinfo {author} {\bibfnamefont {M.}~\bibnamefont {Pita-Vidal}}, \bibinfo {author} {\bibfnamefont {R.}~\bibnamefont {\ifmmode~\check{Z}\else \v{Z}\fi{}itko}}, \bibinfo {author} {\bibfnamefont {L.~J.}\ \bibnamefont {Splitthoff}}, \bibinfo {author} {\bibfnamefont {L.}~\bibnamefont {Gr\"unhaupt}}, \bibinfo {author} {\bibfnamefont {J.~J.}\ \bibnamefont {Wesdorp}}, \bibinfo {author} {\bibfnamefont {Y.}~\bibnamefont {Liu}}, \bibinfo {author} {\bibfnamefont {L.~P.}\ \bibnamefont {Kouwenhoven}}, \bibinfo {author} {\bibfnamefont {R.}~\bibnamefont {Aguado}}, \bibinfo {author} {\bibfnamefont {C.~K.}\ \bibnamefont {Andersen}}, \bibinfo {author} {\bibfnamefont {A.}~\bibnamefont {Kou}},\ and\ \bibinfo {author} {\bibfnamefont {B.}~\bibnamefont {van Heck}},\ }\bibfield  {title} {\bibinfo {title} {Spectroscopy of spin-split andreev levels in a quantum dot with superconducting leads},\ }\href
  {https://doi.org/10.1103/PhysRevLett.131.097001} {\bibfield  {journal} {\bibinfo  {journal} {Phys. Rev. Lett.}\ }\textbf {\bibinfo {volume} {131}},\ \bibinfo {pages} {097001} (\bibinfo {year} {2023})}\BibitemShut {NoStop}%
\bibitem [{\citenamefont {Pita-Vidal}\ \emph {et~al.}(2023)\citenamefont {Pita-Vidal}, \citenamefont {Bargerbos}, \citenamefont {\ifmmode~\check{Z}\else \v{Z}\fi{}itko}, \citenamefont {Splitthoff}, \citenamefont {Gr\"unhaupt}, \citenamefont {Wesdorp}, \citenamefont {Liu}, \citenamefont {Kouwenhoven}, \citenamefont {Aguado}, \citenamefont {van Heck}, \citenamefont {Kou},\ and\ \citenamefont {Andersen}}]{PitaVidal2023}%
  \BibitemOpen
  \bibfield  {author} {\bibinfo {author} {\bibfnamefont {M.}~\bibnamefont {Pita-Vidal}}, \bibinfo {author} {\bibfnamefont {A.}~\bibnamefont {Bargerbos}}, \bibinfo {author} {\bibfnamefont {R.}~\bibnamefont {\ifmmode~\check{Z}\else \v{Z}\fi{}itko}}, \bibinfo {author} {\bibfnamefont {L.~J.}\ \bibnamefont {Splitthoff}}, \bibinfo {author} {\bibfnamefont {L.}~\bibnamefont {Gr\"unhaupt}}, \bibinfo {author} {\bibfnamefont {J.~J.}\ \bibnamefont {Wesdorp}}, \bibinfo {author} {\bibfnamefont {Y.}~\bibnamefont {Liu}}, \bibinfo {author} {\bibfnamefont {L.~P.}\ \bibnamefont {Kouwenhoven}}, \bibinfo {author} {\bibfnamefont {R.}~\bibnamefont {Aguado}}, \bibinfo {author} {\bibfnamefont {B.}~\bibnamefont {van Heck}}, \bibinfo {author} {\bibfnamefont {A.}~\bibnamefont {Kou}},\ and\ \bibinfo {author} {\bibfnamefont {C.~K.}\ \bibnamefont {Andersen}},\ }\bibfield  {title} {\bibinfo {title} {Direct manipulation of a superconducting spin qubit strongly coupled to a transmon qubit},\ }\href
  {http://dx.doi.org/10.1038/s41567-023-02071-x} {\bibfield  {journal} {\bibinfo  {journal} {Nat. Phys.}\ }\textbf {\bibinfo {volume} {19}},\ \bibinfo {pages} {1110–1115} (\bibinfo {year} {2023})}\BibitemShut {NoStop}%
\bibitem [{\citenamefont {Pita-Vidal}\ \emph {et~al.}(2024)\citenamefont {Pita-Vidal}, \citenamefont {Wesdorp}, \citenamefont {Splitthoff}, \citenamefont {Bargerbos}, \citenamefont {Liu}, \citenamefont {Kouwenhoven},\ and\ \citenamefont {Andersen}}]{PitaVidal2024}%
  \BibitemOpen
  \bibfield  {author} {\bibinfo {author} {\bibfnamefont {M.}~\bibnamefont {Pita-Vidal}}, \bibinfo {author} {\bibfnamefont {J.~J.}\ \bibnamefont {Wesdorp}}, \bibinfo {author} {\bibfnamefont {L.~J.}\ \bibnamefont {Splitthoff}}, \bibinfo {author} {\bibfnamefont {A.}~\bibnamefont {Bargerbos}}, \bibinfo {author} {\bibfnamefont {Y.}~\bibnamefont {Liu}}, \bibinfo {author} {\bibfnamefont {L.~P.}\ \bibnamefont {Kouwenhoven}},\ and\ \bibinfo {author} {\bibfnamefont {C.~K.}\ \bibnamefont {Andersen}},\ }\bibfield  {title} {\bibinfo {title} {Strong tunable coupling between two distant superconducting spin qubits},\ }\href {https://doi.org/10.1038/s41567-024-02497-x} {\bibfield  {journal} {\bibinfo  {journal} {Nat. Phys.}\ }\textbf {\bibinfo {volume} {20}},\ \bibinfo {pages} {1158–1163} (\bibinfo {year} {2024})}\BibitemShut {NoStop}%
\bibitem [{\citenamefont {Pita-Vidal}\ \emph {et~al.}(2025)\citenamefont {Pita-Vidal}, \citenamefont {Wesdorp},\ and\ \citenamefont {Andersen}}]{PitaVidal2025}%
  \BibitemOpen
  \bibfield  {author} {\bibinfo {author} {\bibfnamefont {M.}~\bibnamefont {Pita-Vidal}}, \bibinfo {author} {\bibfnamefont {J.~J.}\ \bibnamefont {Wesdorp}},\ and\ \bibinfo {author} {\bibfnamefont {C.~K.}\ \bibnamefont {Andersen}},\ }\bibfield  {title} {\bibinfo {title} {Blueprint for all-to-all-connected superconducting spin qubits},\ }\href {https://doi.org/10.1103/PRXQuantum.6.010308} {\bibfield  {journal} {\bibinfo  {journal} {PRX Quantum}\ }\textbf {\bibinfo {volume} {6}},\ \bibinfo {pages} {010308} (\bibinfo {year} {2025})}\BibitemShut {NoStop}%
\bibitem [{\citenamefont {Makhlin}\ \emph {et~al.}(2001)\citenamefont {Makhlin}, \citenamefont {Sch\"on},\ and\ \citenamefont {Shnirman}}]{Makhlin2001}%
  \BibitemOpen
  \bibfield  {author} {\bibinfo {author} {\bibfnamefont {Y.}~\bibnamefont {Makhlin}}, \bibinfo {author} {\bibfnamefont {G.}~\bibnamefont {Sch\"on}},\ and\ \bibinfo {author} {\bibfnamefont {A.}~\bibnamefont {Shnirman}},\ }\bibfield  {title} {\bibinfo {title} {Quantum-state engineering with josephson-junction devices},\ }\href {https://doi.org/10.1103/RevModPhys.73.357} {\bibfield  {journal} {\bibinfo  {journal} {Rev. Mod. Phys.}\ }\textbf {\bibinfo {volume} {73}},\ \bibinfo {pages} {357} (\bibinfo {year} {2001})}\BibitemShut {NoStop}%
\bibitem [{\citenamefont {Levy}(2002)}]{Levy2002}%
  \BibitemOpen
  \bibfield  {author} {\bibinfo {author} {\bibfnamefont {J.}~\bibnamefont {Levy}},\ }\bibfield  {title} {\bibinfo {title} {Universal quantum computation with spin-$1/2$ pairs and {Heisenberg} exchange},\ }\href {https://doi.org/10.1103/PhysRevLett.89.147902} {\bibfield  {journal} {\bibinfo  {journal} {Phys. Rev. Lett.}\ }\textbf {\bibinfo {volume} {89}},\ \bibinfo {pages} {147902} (\bibinfo {year} {2002})}\BibitemShut {NoStop}%
\bibitem [{\citenamefont {Petta}\ \emph {et~al.}(2005)\citenamefont {Petta}, \citenamefont {Johnson}, \citenamefont {Taylor}, \citenamefont {Laird}, \citenamefont {Yacoby}, \citenamefont {Lukin}, \citenamefont {Marcus}, \citenamefont {Hanson},\ and\ \citenamefont {Gossard}}]{Petta2005}%
  \BibitemOpen
  \bibfield  {author} {\bibinfo {author} {\bibfnamefont {J.~R.}\ \bibnamefont {Petta}}, \bibinfo {author} {\bibfnamefont {A.~C.}\ \bibnamefont {Johnson}}, \bibinfo {author} {\bibfnamefont {J.~M.}\ \bibnamefont {Taylor}}, \bibinfo {author} {\bibfnamefont {E.~A.}\ \bibnamefont {Laird}}, \bibinfo {author} {\bibfnamefont {A.}~\bibnamefont {Yacoby}}, \bibinfo {author} {\bibfnamefont {M.~D.}\ \bibnamefont {Lukin}}, \bibinfo {author} {\bibfnamefont {C.~M.}\ \bibnamefont {Marcus}}, \bibinfo {author} {\bibfnamefont {M.~P.}\ \bibnamefont {Hanson}},\ and\ \bibinfo {author} {\bibfnamefont {A.~C.}\ \bibnamefont {Gossard}},\ }\bibfield  {title} {\bibinfo {title} {Coherent manipulation of coupled electron spins in semiconductor quantum dots},\ }\href {https://doi.org/10.1126/science.1116955} {\bibfield  {journal} {\bibinfo  {journal} {Science}\ }\textbf {\bibinfo {volume} {309}},\ \bibinfo {pages} {2180} (\bibinfo {year} {2005})}\BibitemShut {NoStop}%
\bibitem [{\citenamefont {Geier}\ \emph {et~al.}(2024)\citenamefont {Geier}, \citenamefont {Souto}, \citenamefont {Schulenborg}, \citenamefont {Asaad}, \citenamefont {Leijnse},\ and\ \citenamefont {Flensberg}}]{Geier2024}%
  \BibitemOpen
  \bibfield  {author} {\bibinfo {author} {\bibfnamefont {M.}~\bibnamefont {Geier}}, \bibinfo {author} {\bibfnamefont {R.~S.}\ \bibnamefont {Souto}}, \bibinfo {author} {\bibfnamefont {J.}~\bibnamefont {Schulenborg}}, \bibinfo {author} {\bibfnamefont {S.}~\bibnamefont {Asaad}}, \bibinfo {author} {\bibfnamefont {M.}~\bibnamefont {Leijnse}},\ and\ \bibinfo {author} {\bibfnamefont {K.}~\bibnamefont {Flensberg}},\ }\bibfield  {title} {\bibinfo {title} {Fermion-parity qubit in a proximitized double quantum dot},\ }\href {https://doi.org/10.1103/PhysRevResearch.6.023281} {\bibfield  {journal} {\bibinfo  {journal} {Phys. Rev. Res.}\ }\textbf {\bibinfo {volume} {6}},\ \bibinfo {pages} {023281} (\bibinfo {year} {2024})}\BibitemShut {NoStop}%
\bibitem [{\citenamefont {Steffensen}\ and\ \citenamefont {Yeyati}(2025)}]{Steffensen2025}%
  \BibitemOpen
  \bibfield  {author} {\bibinfo {author} {\bibfnamefont {G.~O.}\ \bibnamefont {Steffensen}}\ and\ \bibinfo {author} {\bibfnamefont {A.~L.}\ \bibnamefont {Yeyati}},\ }\bibfield  {title} {\bibinfo {title} {Yu-shiba-rusinov--bond qubit in a double quantum dot with circuit-qed operation},\ }\href {https://doi.org/10.1103/PRXQuantum.6.020329} {\bibfield  {journal} {\bibinfo  {journal} {PRX Quantum}\ }\textbf {\bibinfo {volume} {6}},\ \bibinfo {pages} {020329} (\bibinfo {year} {2025})}\BibitemShut {NoStop}%
\bibitem [{\citenamefont {Choi}\ \emph {et~al.}(2000)\citenamefont {Choi}, \citenamefont {Bruder},\ and\ \citenamefont {Loss}}]{Choi2000}%
  \BibitemOpen
  \bibfield  {author} {\bibinfo {author} {\bibfnamefont {M.-S.}\ \bibnamefont {Choi}}, \bibinfo {author} {\bibfnamefont {C.}~\bibnamefont {Bruder}},\ and\ \bibinfo {author} {\bibfnamefont {D.}~\bibnamefont {Loss}},\ }\bibfield  {title} {\bibinfo {title} {Spin-dependent {Josephson} current through double quantum dots and measurement of entangled electron states},\ }\href {https://doi.org/10.1103/PhysRevB.62.13569} {\bibfield  {journal} {\bibinfo  {journal} {Phys. Rev. B}\ }\textbf {\bibinfo {volume} {62}},\ \bibinfo {pages} {13569} (\bibinfo {year} {2000})}\BibitemShut {NoStop}%
\bibitem [{\citenamefont {Choi}\ \emph {et~al.}(2001)\citenamefont {Choi}, \citenamefont {Bruder},\ and\ \citenamefont {Loss}}]{Choi2001}%
  \BibitemOpen
  \bibfield  {author} {\bibinfo {author} {\bibfnamefont {M.-S.}\ \bibnamefont {Choi}}, \bibinfo {author} {\bibfnamefont {C.}~\bibnamefont {Bruder}},\ and\ \bibinfo {author} {\bibfnamefont {D.}~\bibnamefont {Loss}},\ }\bibfield  {title} {\bibinfo {title} {Superconductors, quantum dots, and spin entanglement},\ }in\ \href {https://doi.org/10.1007/3-540-45532-9_3} {\emph {\bibinfo {booktitle} {Interact. Electrons Nanostructures. Lect. Notes Physics, vol 579}}},\ \bibinfo {series and number} {\bibinfo {number} {May 2014}},\ \bibinfo {editor} {edited by\ \bibinfo {editor} {\bibfnamefont {R.}~\bibnamefont {Haug}}\ and\ \bibinfo {editor} {\bibfnamefont {H.}~\bibnamefont {Schoeller}}}\ (\bibinfo  {publisher} {Springer, Berlin, Heidelberg},\ \bibinfo {year} {2001})\ pp.\ \bibinfo {pages} {46--66}\BibitemShut {NoStop}%
\bibitem [{\citenamefont {Scher{\"u}bl}\ \emph {et~al.}(2019)\citenamefont {Scher{\"u}bl}, \citenamefont {P{\'a}lyi},\ and\ \citenamefont {Csonka}}]{Scherubl2019}%
  \BibitemOpen
  \bibfield  {author} {\bibinfo {author} {\bibfnamefont {Z.}~\bibnamefont {Scher{\"u}bl}}, \bibinfo {author} {\bibfnamefont {A.}~\bibnamefont {P{\'a}lyi}},\ and\ \bibinfo {author} {\bibfnamefont {S.}~\bibnamefont {Csonka}},\ }\bibfield  {title} {\bibinfo {title} {Transport signatures of an {Andreev} molecule in a quantum dot--superconductor--quantum dot setup},\ }\href {https://doi.org/10.3762/bjnano.10.36} {\bibfield  {journal} {\bibinfo  {journal} {Beilstein J Nanotechnol.}\ }\textbf {\bibinfo {volume} {10}},\ \bibinfo {pages} {363} (\bibinfo {year} {2019})}\BibitemShut {NoStop}%
\bibitem [{\citenamefont {Spethmann}\ \emph {et~al.}(2024)\citenamefont {Spethmann}, \citenamefont {Bosco}, \citenamefont {Hofmann}, \citenamefont {Klinovaja},\ and\ \citenamefont {Loss}}]{Spethmann2024}%
  \BibitemOpen
  \bibfield  {author} {\bibinfo {author} {\bibfnamefont {M.}~\bibnamefont {Spethmann}}, \bibinfo {author} {\bibfnamefont {S.}~\bibnamefont {Bosco}}, \bibinfo {author} {\bibfnamefont {A.}~\bibnamefont {Hofmann}}, \bibinfo {author} {\bibfnamefont {J.}~\bibnamefont {Klinovaja}},\ and\ \bibinfo {author} {\bibfnamefont {D.}~\bibnamefont {Loss}},\ }\bibfield  {title} {\bibinfo {title} {High-fidelity two-qubit gates of hybrid superconducting-semiconducting singlet-triplet qubits},\ }\href {https://doi.org/10.1103/PhysRevB.109.085303} {\bibfield  {journal} {\bibinfo  {journal} {Phys. Rev. B}\ }\textbf {\bibinfo {volume} {109}},\ \bibinfo {pages} {085303} (\bibinfo {year} {2024})}\BibitemShut {NoStop}%
\bibitem [{\citenamefont {Hofstetter}\ \emph {et~al.}(2009)\citenamefont {Hofstetter}, \citenamefont {Csonka}, \citenamefont {Nyg{\aa}rd},\ and\ \citenamefont {Sch{\"o}nenberger}}]{Hofstetter2009}%
  \BibitemOpen
  \bibfield  {author} {\bibinfo {author} {\bibfnamefont {L.}~\bibnamefont {Hofstetter}}, \bibinfo {author} {\bibfnamefont {S.}~\bibnamefont {Csonka}}, \bibinfo {author} {\bibfnamefont {J.}~\bibnamefont {Nyg{\aa}rd}},\ and\ \bibinfo {author} {\bibfnamefont {C.}~\bibnamefont {Sch{\"o}nenberger}},\ }\bibfield  {title} {\bibinfo {title} {Cooper pair splitter realized in a two-quantum-dot y-junction},\ }\href {https://doi.org/https://doi.org/10.1038/nature08432} {\bibfield  {journal} {\bibinfo  {journal} {Nature}\ }\textbf {\bibinfo {volume} {461}},\ \bibinfo {pages} {960} (\bibinfo {year} {2009})}\BibitemShut {NoStop}%
\bibitem [{\citenamefont {Baba}\ \emph {et~al.}(2018)\citenamefont {Baba}, \citenamefont {Jünger}, \citenamefont {Matsuo}, \citenamefont {Baumgartner}, \citenamefont {Sato}, \citenamefont {Kamata}, \citenamefont {Li}, \citenamefont {Jeppesen}, \citenamefont {Samuelson}, \citenamefont {Xu}, \citenamefont {Schönenberger},\ and\ \citenamefont {Tarucha}}]{Baba2018}%
  \BibitemOpen
  \bibfield  {author} {\bibinfo {author} {\bibfnamefont {S.}~\bibnamefont {Baba}}, \bibinfo {author} {\bibfnamefont {C.}~\bibnamefont {Jünger}}, \bibinfo {author} {\bibfnamefont {S.}~\bibnamefont {Matsuo}}, \bibinfo {author} {\bibfnamefont {A.}~\bibnamefont {Baumgartner}}, \bibinfo {author} {\bibfnamefont {Y.}~\bibnamefont {Sato}}, \bibinfo {author} {\bibfnamefont {H.}~\bibnamefont {Kamata}}, \bibinfo {author} {\bibfnamefont {K.}~\bibnamefont {Li}}, \bibinfo {author} {\bibfnamefont {S.}~\bibnamefont {Jeppesen}}, \bibinfo {author} {\bibfnamefont {L.}~\bibnamefont {Samuelson}}, \bibinfo {author} {\bibfnamefont {H.~Q.}\ \bibnamefont {Xu}}, \bibinfo {author} {\bibfnamefont {C.}~\bibnamefont {Schönenberger}},\ and\ \bibinfo {author} {\bibfnamefont {S.}~\bibnamefont {Tarucha}},\ }\bibfield  {title} {\bibinfo {title} {Cooper-pair splitting in two parallel inas nanowires},\ }\href {https://doi.org/10.1088/1367-2630/aac74e} {\bibfield  {journal} {\bibinfo  {journal} {New J. Phys.}\ }\textbf {\bibinfo {volume}
  {20}},\ \bibinfo {pages} {063021} (\bibinfo {year} {2018})}\BibitemShut {NoStop}%
\bibitem [{\citenamefont {Ranni}\ \emph {et~al.}(2021)\citenamefont {Ranni}, \citenamefont {Brange}, \citenamefont {Mannila}, \citenamefont {Flindt},\ and\ \citenamefont {Maisi}}]{Ranni2021}%
  \BibitemOpen
  \bibfield  {author} {\bibinfo {author} {\bibfnamefont {A.}~\bibnamefont {Ranni}}, \bibinfo {author} {\bibfnamefont {F.}~\bibnamefont {Brange}}, \bibinfo {author} {\bibfnamefont {E.~T.}\ \bibnamefont {Mannila}}, \bibinfo {author} {\bibfnamefont {C.}~\bibnamefont {Flindt}},\ and\ \bibinfo {author} {\bibfnamefont {V.~F.}\ \bibnamefont {Maisi}},\ }\bibfield  {title} {\bibinfo {title} {Real-time observation of cooper pair splitting showing strong non-local correlations},\ }\href {https://doi.org/https://doi.org/10.1038/s41467-021-26627-8} {\bibfield  {journal} {\bibinfo  {journal} {Nat. Commun.}\ }\textbf {\bibinfo {volume} {12}},\ \bibinfo {pages} {6358} (\bibinfo {year} {2021})}\BibitemShut {NoStop}%
\bibitem [{\citenamefont {K{\"u}rt{\"o}ssy}\ \emph {et~al.}(2022)\citenamefont {K{\"u}rt{\"o}ssy}, \citenamefont {Scher{\"u}bl}, \citenamefont {F{\"u}l{\"o}p}, \citenamefont {Luk{\'a}cs}, \citenamefont {Kanne}, \citenamefont {Nyg{\aa}rd}, \citenamefont {Makk},\ and\ \citenamefont {Csonka}}]{Kurtossy2022}%
  \BibitemOpen
  \bibfield  {author} {\bibinfo {author} {\bibfnamefont {O.}~\bibnamefont {K{\"u}rt{\"o}ssy}}, \bibinfo {author} {\bibfnamefont {Z.}~\bibnamefont {Scher{\"u}bl}}, \bibinfo {author} {\bibfnamefont {G.}~\bibnamefont {F{\"u}l{\"o}p}}, \bibinfo {author} {\bibfnamefont {I.~E.}\ \bibnamefont {Luk{\'a}cs}}, \bibinfo {author} {\bibfnamefont {T.}~\bibnamefont {Kanne}}, \bibinfo {author} {\bibfnamefont {J.}~\bibnamefont {Nyg{\aa}rd}}, \bibinfo {author} {\bibfnamefont {P.}~\bibnamefont {Makk}},\ and\ \bibinfo {author} {\bibfnamefont {S.}~\bibnamefont {Csonka}},\ }\bibfield  {title} {\bibinfo {title} {Parallel inas nanowires for cooper pair splitters with coulomb repulsion},\ }\href {https://doi.org/https://doi.org/10.1038/s41535-022-00497-9} {\bibfield  {journal} {\bibinfo  {journal} {npj Quantum Mater.}\ }\textbf {\bibinfo {volume} {7}},\ \bibinfo {pages} {88} (\bibinfo {year} {2022})}\BibitemShut {NoStop}%
\bibitem [{\citenamefont {Recher}\ \emph {et~al.}(2001)\citenamefont {Recher}, \citenamefont {Sukhorukov},\ and\ \citenamefont {Loss}}]{Recher2001}%
  \BibitemOpen
  \bibfield  {author} {\bibinfo {author} {\bibfnamefont {P.}~\bibnamefont {Recher}}, \bibinfo {author} {\bibfnamefont {E.~V.}\ \bibnamefont {Sukhorukov}},\ and\ \bibinfo {author} {\bibfnamefont {D.}~\bibnamefont {Loss}},\ }\bibfield  {title} {\bibinfo {title} {Andreev tunneling, coulomb blockade, and resonant transport of nonlocal spin-entangled electrons},\ }\href {https://doi.org/10.1103/PhysRevB.63.165314} {\bibfield  {journal} {\bibinfo  {journal} {Phys. Rev. B}\ }\textbf {\bibinfo {volume} {63}},\ \bibinfo {pages} {165314} (\bibinfo {year} {2001})}\BibitemShut {NoStop}%
\bibitem [{\citenamefont {J{\"u}nger}\ \emph {et~al.}(2023)\citenamefont {J{\"u}nger}, \citenamefont {Lehmann}, \citenamefont {Dick}, \citenamefont {Thelander}, \citenamefont {Sch{\"o}nenberger},\ and\ \citenamefont {Baumgartner}}]{Junger2023}%
  \BibitemOpen
  \bibfield  {author} {\bibinfo {author} {\bibfnamefont {C.}~\bibnamefont {J{\"u}nger}}, \bibinfo {author} {\bibfnamefont {S.}~\bibnamefont {Lehmann}}, \bibinfo {author} {\bibfnamefont {K.~A.}\ \bibnamefont {Dick}}, \bibinfo {author} {\bibfnamefont {C.}~\bibnamefont {Thelander}}, \bibinfo {author} {\bibfnamefont {C.}~\bibnamefont {Sch{\"o}nenberger}},\ and\ \bibinfo {author} {\bibfnamefont {A.}~\bibnamefont {Baumgartner}},\ }\bibfield  {title} {\bibinfo {title} {Intermediate states in andreev bound state fusion},\ }\href {https://doi.org/https://doi.org/10.1038/s42005-023-01273-2} {\bibfield  {journal} {\bibinfo  {journal} {Commun. Phys.}\ }\textbf {\bibinfo {volume} {6}},\ \bibinfo {pages} {190} (\bibinfo {year} {2023})}\BibitemShut {NoStop}%
\bibitem [{\citenamefont {K{\"u}rt{\"o}ssy}\ \emph {et~al.}(2021)\citenamefont {K{\"u}rt{\"o}ssy}, \citenamefont {Scher{\"u}bl}, \citenamefont {F{\"u}l{\"o}p}, \citenamefont {Lukács}, \citenamefont {Kanne}, \citenamefont {Nyg{\aa}rd}, \citenamefont {Makk},\ and\ \citenamefont {Csonka}}]{Kurtossy2021}%
  \BibitemOpen
  \bibfield  {author} {\bibinfo {author} {\bibfnamefont {O.}~\bibnamefont {K{\"u}rt{\"o}ssy}}, \bibinfo {author} {\bibfnamefont {Z.}~\bibnamefont {Scher{\"u}bl}}, \bibinfo {author} {\bibfnamefont {G.}~\bibnamefont {F{\"u}l{\"o}p}}, \bibinfo {author} {\bibfnamefont {I.~E.}\ \bibnamefont {Lukács}}, \bibinfo {author} {\bibfnamefont {T.}~\bibnamefont {Kanne}}, \bibinfo {author} {\bibfnamefont {J.}~\bibnamefont {Nyg{\aa}rd}}, \bibinfo {author} {\bibfnamefont {P.}~\bibnamefont {Makk}},\ and\ \bibinfo {author} {\bibfnamefont {S.}~\bibnamefont {Csonka}},\ }\bibfield  {title} {\bibinfo {title} {Andreev molecule in parallel inas nanowires},\ }\href {https://doi.org/10.1021/acs.nanolett.1c01956} {\bibfield  {journal} {\bibinfo  {journal} {Nano Letters}\ }\textbf {\bibinfo {volume} {21}},\ \bibinfo {pages} {7929} (\bibinfo {year} {2021})}\BibitemShut {NoStop}%
\bibitem [{\citenamefont {Leijnse}\ and\ \citenamefont {Flensberg}(2013)}]{Leijnse2013}%
  \BibitemOpen
  \bibfield  {author} {\bibinfo {author} {\bibfnamefont {M.}~\bibnamefont {Leijnse}}\ and\ \bibinfo {author} {\bibfnamefont {K.}~\bibnamefont {Flensberg}},\ }\bibfield  {title} {\bibinfo {title} {Coupling spin qubits via superconductors},\ }\href {https://doi.org/10.1103/PhysRevLett.111.060501} {\bibfield  {journal} {\bibinfo  {journal} {Phys. Rev. Lett.}\ }\textbf {\bibinfo {volume} {111}},\ \bibinfo {pages} {060501} (\bibinfo {year} {2013})}\BibitemShut {NoStop}%
\bibitem [{\citenamefont {Matsuo}\ \emph {et~al.}(2022)\citenamefont {Matsuo}, \citenamefont {Lee}, \citenamefont {Chang}, \citenamefont {Sato}, \citenamefont {Ueda}, \citenamefont {Palmstr{\o}m},\ and\ \citenamefont {Tarucha}}]{Matsuo2022}%
  \BibitemOpen
  \bibfield  {author} {\bibinfo {author} {\bibfnamefont {S.}~\bibnamefont {Matsuo}}, \bibinfo {author} {\bibfnamefont {J.~S.}\ \bibnamefont {Lee}}, \bibinfo {author} {\bibfnamefont {C.-Y.}\ \bibnamefont {Chang}}, \bibinfo {author} {\bibfnamefont {Y.}~\bibnamefont {Sato}}, \bibinfo {author} {\bibfnamefont {K.}~\bibnamefont {Ueda}}, \bibinfo {author} {\bibfnamefont {C.~J.}\ \bibnamefont {Palmstr{\o}m}},\ and\ \bibinfo {author} {\bibfnamefont {S.}~\bibnamefont {Tarucha}},\ }\bibfield  {title} {\bibinfo {title} {Observation of nonlocal josephson effect on double inas nanowires},\ }\href {https://doi.org/https://doi.org/10.1038/s42005-022-00994-0} {\bibfield  {journal} {\bibinfo  {journal} {Commun. Phys.}\ }\textbf {\bibinfo {volume} {5}},\ \bibinfo {pages} {221} (\bibinfo {year} {2022})}\BibitemShut {NoStop}%
\bibitem [{\citenamefont {Ueda}\ \emph {et~al.}(2019)\citenamefont {Ueda}, \citenamefont {Matsuo}, \citenamefont {Kamata}, \citenamefont {Baba}, \citenamefont {Sato}, \citenamefont {Takeshige}, \citenamefont {Li}, \citenamefont {Jeppesen}, \citenamefont {Samuelson}, \citenamefont {Xu},\ and\ \citenamefont {Tarucha}}]{Ueda2019}%
  \BibitemOpen
  \bibfield  {author} {\bibinfo {author} {\bibfnamefont {K.}~\bibnamefont {Ueda}}, \bibinfo {author} {\bibfnamefont {S.}~\bibnamefont {Matsuo}}, \bibinfo {author} {\bibfnamefont {H.}~\bibnamefont {Kamata}}, \bibinfo {author} {\bibfnamefont {S.}~\bibnamefont {Baba}}, \bibinfo {author} {\bibfnamefont {Y.}~\bibnamefont {Sato}}, \bibinfo {author} {\bibfnamefont {Y.}~\bibnamefont {Takeshige}}, \bibinfo {author} {\bibfnamefont {K.}~\bibnamefont {Li}}, \bibinfo {author} {\bibfnamefont {S.}~\bibnamefont {Jeppesen}}, \bibinfo {author} {\bibfnamefont {L.}~\bibnamefont {Samuelson}}, \bibinfo {author} {\bibfnamefont {H.}~\bibnamefont {Xu}},\ and\ \bibinfo {author} {\bibfnamefont {S.}~\bibnamefont {Tarucha}},\ }\bibfield  {title} {\bibinfo {title} {Dominant nonlocal superconducting proximity effect due to electron-electron interaction in a ballistic double nanowire},\ }\href {https://doi.org/10.1126/sciadv.aaw2194} {\bibfield  {journal} {\bibinfo  {journal} {Sci. Adv.}\ }\textbf {\bibinfo {volume} {5}},\ \bibinfo {pages}
  {eaaw2194} (\bibinfo {year} {2019})}\BibitemShut {NoStop}%
\bibitem [{\citenamefont {Kanne}\ \emph {et~al.}(2022)\citenamefont {Kanne}, \citenamefont {Olsteins}, \citenamefont {Marnauza}, \citenamefont {Vekris}, \citenamefont {Estrada~Salda{\~n}a}, \citenamefont {Lori\c`}, \citenamefont {Schlosser}, \citenamefont {Ross}, \citenamefont {Csonka}, \citenamefont {Grove-Rasmussen},\ and\ \citenamefont {Nyg{\aa}rd}}]{Kanne2022}%
  \BibitemOpen
  \bibfield  {author} {\bibinfo {author} {\bibfnamefont {T.}~\bibnamefont {Kanne}}, \bibinfo {author} {\bibfnamefont {D.}~\bibnamefont {Olsteins}}, \bibinfo {author} {\bibfnamefont {M.}~\bibnamefont {Marnauza}}, \bibinfo {author} {\bibfnamefont {A.}~\bibnamefont {Vekris}}, \bibinfo {author} {\bibfnamefont {J.~C.}\ \bibnamefont {Estrada~Salda{\~n}a}}, \bibinfo {author} {\bibfnamefont {S.}~\bibnamefont {Lori\c`}}, \bibinfo {author} {\bibfnamefont {R.~D.}\ \bibnamefont {Schlosser}}, \bibinfo {author} {\bibfnamefont {D.}~\bibnamefont {Ross}}, \bibinfo {author} {\bibfnamefont {S.}~\bibnamefont {Csonka}}, \bibinfo {author} {\bibfnamefont {K.}~\bibnamefont {Grove-Rasmussen}},\ and\ \bibinfo {author} {\bibfnamefont {J.}~\bibnamefont {Nyg{\aa}rd}},\ }\bibfield  {title} {\bibinfo {title} {Double nanowires for hybrid quantum devices},\ }\href {https://doi.org/https://doi.org/10.1002/adfm.202107926} {\bibfield  {journal} {\bibinfo  {journal} {Adv. Funct. Mater.}\ }\textbf {\bibinfo {volume} {32}},\ \bibinfo {pages}
  {2107926} (\bibinfo {year} {2022})}\BibitemShut {NoStop}%
\bibitem [{Note1()}]{Note1}%
  \BibitemOpen
  \bibinfo {note} {Choi et al. \cite {Choi2000} calculated the result in Eq.~\protect \eqref {eq:H_DDJJ} for $\Delta \ll \protect \mathcal {U}$ and without Zeeman field. However, their result stays valid without these restrictions, thus, for $\Gamma , h_{L/R}\ll |\epsilon |\ll \protect \mathcal {U}$ and $\Gamma ,h_{L/R}\ll \Delta $. The reason is that (i) a larger $\Delta $ does not enable new tunnel paths and (ii) Zeeman fields only appear in the energy denominators of perturbation theory, where they are much smaller than the excitation energies $\protect \mathcal {O}(\Delta , |\epsilon |)$.}\BibitemShut {Stop}%
\bibitem [{\citenamefont {Jirovec}\ \emph {et~al.}(2021)\citenamefont {Jirovec}, \citenamefont {Hofmann}, \citenamefont {Ballabio}, \citenamefont {Mutter}, \citenamefont {Tavani}, \citenamefont {Botifoll}, \citenamefont {Crippa}, \citenamefont {Kukucka}, \citenamefont {Sagi}, \citenamefont {Martins}, \citenamefont {Saez-Mollejo}, \citenamefont {Prieto}, \citenamefont {Borovkov}, \citenamefont {Arbiol}, \citenamefont {Chrastina}, \citenamefont {Isella},\ and\ \citenamefont {Katsaros}}]{Jirovec2021}%
  \BibitemOpen
  \bibfield  {author} {\bibinfo {author} {\bibfnamefont {D.}~\bibnamefont {Jirovec}}, \bibinfo {author} {\bibfnamefont {A.}~\bibnamefont {Hofmann}}, \bibinfo {author} {\bibfnamefont {A.}~\bibnamefont {Ballabio}}, \bibinfo {author} {\bibfnamefont {P.~M.}\ \bibnamefont {Mutter}}, \bibinfo {author} {\bibfnamefont {G.}~\bibnamefont {Tavani}}, \bibinfo {author} {\bibfnamefont {M.}~\bibnamefont {Botifoll}}, \bibinfo {author} {\bibfnamefont {A.}~\bibnamefont {Crippa}}, \bibinfo {author} {\bibfnamefont {J.}~\bibnamefont {Kukucka}}, \bibinfo {author} {\bibfnamefont {O.}~\bibnamefont {Sagi}}, \bibinfo {author} {\bibfnamefont {F.}~\bibnamefont {Martins}}, \bibinfo {author} {\bibfnamefont {J.}~\bibnamefont {Saez-Mollejo}}, \bibinfo {author} {\bibfnamefont {I.}~\bibnamefont {Prieto}}, \bibinfo {author} {\bibfnamefont {M.}~\bibnamefont {Borovkov}}, \bibinfo {author} {\bibfnamefont {J.}~\bibnamefont {Arbiol}}, \bibinfo {author} {\bibfnamefont {D.}~\bibnamefont {Chrastina}}, \bibinfo {author} {\bibfnamefont {G.}~\bibnamefont
  {Isella}},\ and\ \bibinfo {author} {\bibfnamefont {G.}~\bibnamefont {Katsaros}},\ }\bibfield  {title} {\bibinfo {title} {A singlet-triplet hole spin qubit in planar {{Ge}}},\ }\href {https://doi.org/10.1038/s41563-021-01022-2} {\bibfield  {journal} {\bibinfo  {journal} {Nat. Mater.}\ }\textbf {\bibinfo {volume} {20}},\ \bibinfo {pages} {1106–1112} (\bibinfo {year} {2021})}\BibitemShut {NoStop}%
\bibitem [{Note2()}]{Note2}%
  \BibitemOpen
  \bibinfo {note} {Setting the superconducting phase difference to $\phi =0$ corresponds to the X flux set point described in Sec. \ref {sec:qubit_gates}.}\BibitemShut {Stop}%
\bibitem [{Note3()}]{Note3}%
  \BibitemOpen
  \bibinfo {note} {The $e$ is the elementary charge.}\BibitemShut {Stop}%
\bibitem [{Note4()}]{Note4}%
  \BibitemOpen
  \bibinfo {note} {The $\hbar =\protect \frac {h}{2\pi }$ is the reduced Planck's constant.}\BibitemShut {Stop}%
\bibitem [{Note5()}]{Note5}%
  \BibitemOpen
  \bibinfo {note} {We dropped the constant $-\protect \frac {1}{2}\DOTSB \sum@ \slimits@ _{j=1}^NE_\protect \text {CA}^j$.}\BibitemShut {Stop}%
\bibitem [{Note6()}]{Note6}%
  \BibitemOpen
  \bibinfo {note} {Moving into a rotated frame, given by the transformation $U_R(t)=e ^{-i\delta h\tau _j^zt/(2\hbar )}$, transforms $\tau _x$ in $H_j^{1q}$ into $\left [\cos (\delta h t/\hbar )\tau _j^x-\sin (\delta h t/\hbar )\tau _j^y\right ]$. Therefore, a flux pulse applied at time $t=\protect \frac {\pi \hbar N_t}{|\delta h|}$, $N_t \in \protect \mathbb {N}$ enables an $x$ rotation, and a flux pulse at $t=\protect \frac {\pi \hbar (N_t+1/2)}{|\delta h|}$ enables a $y$ rotation.}\BibitemShut {Stop}%
\bibitem [{Note7()}]{Note7}%
  \BibitemOpen
  \bibinfo {note} {For example, we get a CZ gate with $CZ = e^{i\pi /4} H_j H_k R_k(j) R_j(k) e^{-i T H^\protect \text {INT}_{jk} / \hbar } H_j H_k$. Here, $H_j$ is a Hadamard gate on qubit $j$ and $R_k(j)= \exp \left (i\protect \frac {\pi }{4}\tau _j^x\left [\protect \frac {2E_J+E_\protect \text {CA}^{j}}{E_\protect \text {CA}^{k}}\right ]\right )$ is a rotation around the $x$ axis of qubit $j$.}\BibitemShut {Stop}%
\bibitem [{Note8()}]{Note8}%
  \BibitemOpen
  \bibinfo {note} {The calculation of Makhlin invariants \cite {Makhlin2002} may be helpful in finding the exact gate sequence.}\BibitemShut {Stop}%
\bibitem [{\citenamefont {Krantz}\ \emph {et~al.}(2019)\citenamefont {Krantz}, \citenamefont {Kjaergaard}, \citenamefont {Yan}, \citenamefont {Orlando}, \citenamefont {Gustavsson},\ and\ \citenamefont {Oliver}}]{Krantz2019}%
  \BibitemOpen
  \bibfield  {author} {\bibinfo {author} {\bibfnamefont {P.}~\bibnamefont {Krantz}}, \bibinfo {author} {\bibfnamefont {M.}~\bibnamefont {Kjaergaard}}, \bibinfo {author} {\bibfnamefont {F.}~\bibnamefont {Yan}}, \bibinfo {author} {\bibfnamefont {T.~P.}\ \bibnamefont {Orlando}}, \bibinfo {author} {\bibfnamefont {S.}~\bibnamefont {Gustavsson}},\ and\ \bibinfo {author} {\bibfnamefont {W.~D.}\ \bibnamefont {Oliver}},\ }\bibfield  {title} {\bibinfo {title} {A {{Quantum Engineer}}'s {{Guide}} to {{Superconducting Qubits}}},\ }\href@noop {} {\bibfield  {journal} {\bibinfo  {journal} {Applied Physics Reviews}\ }\textbf {\bibinfo {volume} {6}},\ \bibinfo {pages} {021318} (\bibinfo {year} {2019})}\BibitemShut {NoStop}%
\bibitem [{Note9()}]{Note9}%
  \BibitemOpen
  \bibinfo {note} {In that case, the Hamiltonian describing one qubit simply becomes $H_\protect \text {DDJJ}'=\protect \frac {\delta h}{2}\tau _j^z$ (for $\Gamma =0$), see App.~\ref {Sec:general_H_largeGamma}.}\BibitemShut {Stop}%
\bibitem [{\citenamefont {Winkler}(2003)}]{Winkler2003}%
  \BibitemOpen
  \bibfield  {author} {\bibinfo {author} {\bibfnamefont {R.}~\bibnamefont {Winkler}},\ }\href {https://www.physics.udel.edu/~bnikolic/QTTG/NOTES/SPINTRONICS/WINKLER=spin_orbit_coupling_effect_in_2d_electron_and_hole_systems.pdf} {\emph {\bibinfo {title} {{Spin-Orbit Coupling Effects in Two-Dimensional Electron and Hole Systems, Springer Tracts in Modern Physics 191, VII-IX}}}}\ (\bibinfo  {publisher} {Springer-Verlag, Berlin Heidelberg},\ \bibinfo {year} {2003})\BibitemShut {NoStop}%
\bibitem [{Note10()}]{Note10}%
  \BibitemOpen
  \bibinfo {note} {We assume that the density of states in the superconductors is constant, $\rho (\epsilon )=\rho _F$, and in some calculations we take the principal value of the integral.}\BibitemShut {Stop}%
\bibitem [{Note11()}]{Note11}%
  \BibitemOpen
  \bibinfo {note} {The coupling energy is $\protect \frac {-\Gamma _1-\Gamma _2}{4\Delta }(\DOTSB \sum@ \slimits@ _{\alpha }\epsilon _{\alpha }+\sigma \protect \frac {h_{\alpha }}{2})$.}\BibitemShut {Stop}%
\bibitem [{Note12()}]{Note12}%
  \BibitemOpen
  \bibinfo {note} {The superconductors renormalize the dot energies $\epsilon _{L,R}$ and Zeeman energies $h_{L,R}$ of the states, so from now on, the energies $\epsilon _\alpha $ and $h_{\alpha }$ refer to the corresponding normalized energies. Here, we assumed that the density of states in the leads is symmetric around the Fermi energy.}\BibitemShut {Stop}%
\bibitem [{Note13()}]{Note13}%
  \BibitemOpen
  \bibinfo {note} {We dropped a spin-independent constant of $-\protect \frac {E_C}{4}$}\BibitemShut {NoStop}%
\bibitem [{\citenamefont {Walls}\ and\ \citenamefont {Milburn}(2025)}]{walls_quantum_2025}%
  \BibitemOpen
  \bibfield  {author} {\bibinfo {author} {\bibfnamefont {D.~F.}\ \bibnamefont {Walls}}\ and\ \bibinfo {author} {\bibfnamefont {G.~J.}\ \bibnamefont {Milburn}},\ }\href {https://doi.org/10.1007/978-3-031-84177-4} {\emph {\bibinfo {title} {Quantum {{Optics}}}}},\ Graduate Texts in Physics\ (\bibinfo  {publisher} {Springer Nature Switzerland},\ \bibinfo {address} {Cham},\ \bibinfo {year} {2025})\BibitemShut {NoStop}%
\bibitem [{\citenamefont {Makhlin}(2002)}]{Makhlin2002}%
  \BibitemOpen
  \bibfield  {author} {\bibinfo {author} {\bibfnamefont {Y.}~\bibnamefont {Makhlin}},\ }\bibfield  {title} {\bibinfo {title} {Nonlocal properties of two-qubit gates and mixed states, and the optimization of quantum computations},\ }\href {https://doi.org/https://doi.org/10.1023/A:1022144002391} {\bibfield  {journal} {\bibinfo  {journal} {Quantum Inf. Process.}\ }\textbf {\bibinfo {volume} {1}},\ \bibinfo {pages} {243} (\bibinfo {year} {2002})}\BibitemShut {NoStop}%
\end{thebibliography}%

\end{document}